\documentclass[%
superscriptaddress,
 amsmath,amssymb,
prl,
twocolumn,
titlepage
]{revtex4-2}

\usepackage{graphicx}
\usepackage{dcolumn}
\usepackage{bm}
\usepackage[colorlinks=true,linkcolor=blue,urlcolor=blue,citecolor=blue]{hyperref}
\usepackage{bbold}

\usepackage{color}

\newcommand{\dropcap}[1]{#1}

\newcommand\sbfig[1]{(#1)}

\newcommand{\beginsupplement}{%
        \setcounter{table}{0}
        \renewcommand{\thetable}{S\arabic{table}}%
        \setcounter{figure}{0}
        \renewcommand{\thefigure}{S\arabic{figure}}%
        \setcounter{equation}{0}
        \def\theequation{S\arabic{equation}}
     }

\renewcommand{\vec}{\mathbf}

\begin{document}


\title{Inverse Design of Discrete Mechanical Metamaterials}

\author{Henrik Ronellenfitsch}
 \affiliation{Department of Mathematics, Massachusetts Institute of Technology, 77 Massachusetts Avenue, Cambridge, MA 02139-4307, U.S.A.}
\author{Norbert Stoop}
 \affiliation{Department of Mathematics, Massachusetts Institute of Technology, 77 Massachusetts Avenue, Cambridge, MA 02139-4307, U.S.A.}
\affiliation{Institute of Building Materials, ETH Z\"urich, Stefano-Franscini-Platz 3, 8093 Z\"urich, Switzerland}

\author{Josephine Yu}
\affiliation{Department of Physics, Massachusetts Institute of Technology, 77 Massachusetts Avenue, Cambridge, MA 02139-4307, U.S.A.}

\author{Aden Forrow}
\affiliation{Department of Mathematics, Massachusetts Institute of Technology, 77 Massachusetts Avenue, Cambridge, MA 02139-4307, U.S.A.}
\affiliation{Mathematical Institute, University of Oxford, Andrew Wiles Building, Radcliffe Observatory Quarter, Woodstock Road, Oxford, OX2 6GG, U.K.}

\author{J\"orn Dunkel}
\affiliation{Department of Mathematics, Massachusetts Institute of Technology, 77 Massachusetts Avenue, Cambridge, MA 02139-4307, U.S.A.}
\date{\today}

\begin{abstract}
Mechanical and phononic metamaterials exhibiting negative elastic moduli, gapped vibrational spectra, or topologically protected modes enable precise control of structural and acoustic functionalities. While much progress has been made in their experimental and theoretical characterization, the inverse design of mechanical metamaterials with arbitrarily programmable spectral properties and mode localization remains an unsolved problem. Here, we present a flexible computational inverse-design framework that allows the efficient tuning of one  or more gaps at nearly arbitrary positions in the spectrum of discrete phononic metamaterial structures.  The underlying algorithm optimizes the linear response of elastic networks directly, is applicable to ordered and disordered structures, scales efficiently in 2D and 3D, and can be combined with a wide range of numerical optimization schemes. We illustrate the broad practical potential of this approach by designing mechanical bandgap switches that open and close pre-programmed spectral gaps in response to an externally applied stimulus such as shear or compression. We further show that the designed structures can host topologically protected edge modes, and validate the numerical predictions through explicit 3D finite element simulations of continuum elastica with experimentally relevant material parameters. Generally, this network-based inverse design paradigm offers a direct pathway towards manufacturing phononic metamaterials, DNA origami structures and topolectric circuits that can realize a wide range of static and dynamic target functionalities.
\end{abstract}

\maketitle

\section{Introduction}
\begin{figure*}
\centering
  \includegraphics[width=.92\textwidth]{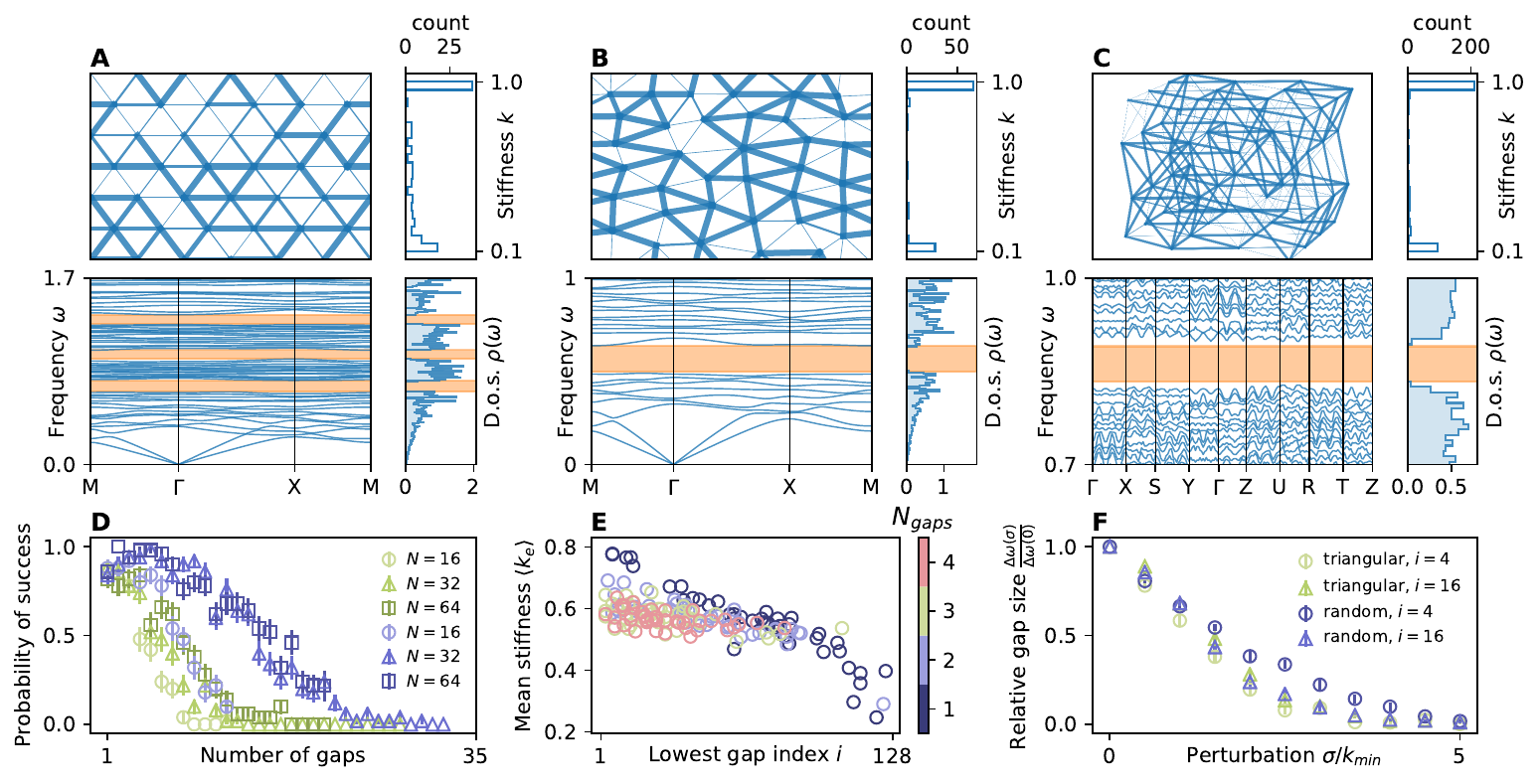}
  \caption{
  \textbf{Designing bandgaps in 2D and 3D phononic networks by linear response optimization (LRO).}
  (\textbf{A})~Triangular $6\times 6$ unit cell of a 2D periodic network with three
  tuned bandgaps (orange). In the optimized network, most springs have stiffness values at the boundaries of the permitted interval $[0.1,1]$. The band structure was computed over the points $\Gamma=(0,0)$, $M=(0,\pi/b)$, $X=(\pi/a, \pi/b)$, where
  $a,b$ are the dimensions of the rectangular unit cell in $x$ and $y$ direction.
  (\textbf{B})~Randomized 2D Delaunay network
  topology with one tuned gap (Movie~1). The bimodal stiffness distribution yields a stiff scaffold network with soft \lq holes\rq{}, realizing impedance mismatch.
  (\textbf{C})~A single gap programmed into a 3D
  tetrahedral network. The band structure was computed over the points for
  the orthorhombic unit cell from Ref.~\cite{Setyawan2010}.
  (\textbf{D})~The probability of successfully tuning a predetermined number of gaps into an $N$-vertex unit cell is significantly higher for   networks  with amorphous topology (blue) than for triangular
  grids (green). Each data point is an average over 50
  random choices for a fixed number of target gaps; error bars indicate standard deviation.
  (\textbf{E})~The mean stiffness is negatively correlated
  with the index of the lowest tuned gap. As the number
  of gaps increases, the stiffness distribution becomes less
  binary and less correlated to gap position. We show data for a $8\times 8$
  triangular grid, but randomized topologies
  behave similarly.
  (\textbf{F})~Gap robustness is independent of gap position
  and network topology. For both triangular and amorphous
  unit cells with a single gap at mode index $i$,
  the addition of zero-mean Gaussian noise with variance $\sigma$ to
  the stiffnesses causes a decrease in the gap size
  as $\sigma \to k_{\mathrm{min}}$.
  \label{fg:figure1}}
\end{figure*}


\dropcap{P}hononic metamaterials~\cite{Cummer2016} offer exciting opportunities to precisely control the passage of sound waves in applications ranging from acoustic cloaking~\cite{Zhang2011} and lensing~\cite{Brunet2014} to art~\cite{Martinez-Sala1995}. Often, the counter-intuitive properties of such materials arise from gaps in their vibrational spectra, which prevent or attenuate the conduction of sound over specific frequency ranges. Recent advances in additive manufacturing techniques such as 3D printing~\cite{Bhattacharjee2016} and lithography~\cite{Buckmann2012}, make it possible now to assemble and experimentally characterize geometrically complex mechanical metamaterials~\cite{Bertoldi2017}. In parallel, insightful theoretical studies~\cite{Ma2016} have substantially improved our understanding of the effects of lattice geometry and broken symmetries on band structure, mode localization and topological protection in phononic metamaterials~\cite{Deymier2013,Susstrunk2016}.  Despite such important progress, major open challenges remain regarding the inverse design of
 mechanical metamaterials with arbitrarily programmable spectra and modes. In particular, the systematic design of highly amorphous, dynamically tunable structures that allow the controlled switching between gapped and ungapped states remains an unsolved problem. Here, we introduce and demonstrate a theoretical and computational framework to solve static and dynamic inverse design tasks for a broad class of discrete mechanical metamaterials.

\par 
The problem of designing ordered and disordered materials with desired spectral properties has a long and rich history~\cite{Sigmund2003,Cummer2016,Ma2016,Deymier2013,Bendsoe2003}.
In the context of modern metamaterials, engineered bandgaps were studied first in optics using approaches ranging from direct shape-optimization in periodic~\cite{Diest2013} and disordered systems~\cite{Rechtsman2008} to the tuning
hyperuniform geometries~\cite{Man2013,Florescu2009} to realize desired electromagnetic absorption and transmission spectra.
Over the past years, the underlying concepts were generalized to acoustic bandgap engineering through the exploitation
of locally resonant units~\cite{Ma2016}, hierarchical self-similar
lattices~\cite{Mousanezhad2015}, gyroscopic materials with
topologically protected modes~\cite{Nash2015,Wang2015},
and topology optimization of continuous materials~\cite{Sigmund2003,Bendsoe2003,Men2014}. The research in these areas has identified resonances~\cite{OlssonIII2009} and impedance mismatch between different material components~\cite{Goffaux2001} as the two primary mechanisms underlying bandgap formation.  In parallel, recent work on quantum~\cite{Agarwala2017} and quantum-like~\cite{Mitchell2018} systems  showed that disorder can promote bandgap formation. Building on these complementary insights,
the inverse design approach described below achieves programmable bandgaps and mode localization by  optimizing the linear physical response, typically yielding highly disordered unit cells.

\par 
Our algorithmic framework is based on a discrete network representation of the underlying mechanical structure, rendering it equally applicable to a broad range of ordered  and disordered~\cite{Yan2017,Reid2017,Flechsig2017,Rocks2017,Goodrich2015,Rocks2019} systems.  Compared with earlier work which focused on the direct numerical tuning of spectral bandgaps in continuum~\cite{Sigmund2003} and  discrete~\cite{Jensen2003} materials through topology optimization,  the indirect response-optimization approach pursued here offers two essential  advantages: First, conceptually it enables an interpretation of the gapped networks as response-minimized metamaterial structures. Second, computationally the scheme can be easily combined with efficient gradient-based methods.
The method is not restricted to the commonly considered low-lying bandgaps in highly symmetric structures, but instead allows the placement of one or more bandgaps at nearly arbitrary positions in the spectrum (Fig.~\ref{fg:figure1}).
We will show how these facts can be used to design phononic switches with prescribed spectral structure under different global deformations (Fig.~\ref{fg:figure2}).  To connect with experiments, we will also demonstrate through 3D finite element (FE) simulations for continuum elastica with realistic material parameters that bond bending alone can suffice to capture and tune the dynamics of continuum elastic networks (Fig.~\ref{fg:figure3}). Our discussion concludes by showing that the response-optimized networks can host protected chiral edge modes, thus enabling the inverse design and precise control of topological metamaterial properties  (Fig.~\ref{fg:figure4}).

\section*{Theory and Results}

\subsection*{Discrete mechanical networks}
The inverse design algorithm described below optimizes the linear response over a set of experimentally tunable system parameters. While the approach generalizes to arbitrary dynamical systems that can be linearized in the neighborhood of fixed points, we focus here on discrete mechanical networks consisting of  $n$ identical point masses $m_v=m$ connected by springs with stiffnesses $\{k_e\}$, where $v$ is a vertex and $e$ is an edge of the network. The elastic energy of the network reads,
\begin{align}
    V = \frac{1}{2} \sum_e k_e \left(\ell_e - \ell_e^{(0)}\right)^2,
    \label{eq:nonlin}
\end{align}
where $k_e$ is the stiffness of spring $e$, $\ell_e$ is its length
and $\ell_e^{(0)}$ is its rest length. In this case, we are interested in programming spectral properties by optimizing over the set of spring constants $\{k_e\}$. To linear order, the dynamics of the network near the equilibrium configuration is given by
\begin{align}
m\,\ddot{\mathbf u} + K \mathbf u = 0,
\label{eq:dynamics}
\end{align}
where $\mathbf u=(u_v)$ is the vector of mass displacements and
$K = Q k Q^\top$ is the
stiffness matrix with $k$ the diagonal matrix of individual
spring stiffnesses
and $Q$ the compatibility matrix encoding the relative geometric relationships
between the masses (Ref.~\cite{Lubensky2015} and Supplementary Material).
Equation~(\ref{eq:dynamics}) can be simplified further  by expanding into
eigenmodes defined by the relation $K\mathbf u_i = m \omega^2_i\, \mathbf u_i$,
where the set $\{\omega_i\}$ constitutes the spectrum of excitation frequencies.
The goal of the spectral optimization is then to construct networks with spring
constants $\{k_e\}$ that realize a desired frequency spectrum $\{\omega_i\}$.
Specifically, to achieve a reduced acoustic response, we would like to place
large gaps between predetermined consecutive eigenvalues $\omega_i$.

\subsection*{Response optimization}
To design one or more spectral gaps at desired locations in the spectrum, we formulate a linear response optimization (LRO) scheme that creates suitable impedance mismatches (Fig.~\ref{fg:figure1}). Unlike brute-force~\cite{Men2010} optimization (Supplementary Material), the LRO  framework  yields a differentiable objective function and is equally applicable to undeformed and deformed networks (Fig.~\ref{fg:figure2}).   The linear response to harmonic forcing $\mathbf{F}e^{i\omega t}$ is given by $\mathbf{u}(t) = e^{i\omega t} G(\omega;k)\mathbf{F}$, where the response function is $G(\omega;k) = (-m\omega^2\mathbb{1} + K)^{-1}$.
The time-averaged covariances  $C_{vv'}(\mathbf{F}) = \langle u_v(t) u_{v'}(t) \rangle_t$ of the vertex responses can then be written in matrix form as
$C(\mathbf{F}) = G(\omega;k) \mathbf{F} \mathbf{F}^H G(\omega;k)^H$, where superscript
$H$ denotes the Hermitian transpose. The strength of each node's response
at frequency $\omega$ is encoded in the diagonal entries. Further averaging over an ensemble of independent,
identically distributed random forcings $\mathbf{F}$ with covariances
$\langle \mathbf{F}\mathbf{F}^H\rangle = \mathbb{1}$
and summing over the diagonal of $\langle C(\mathbf{F})\rangle_{\mathbf{F}}$, we obtain the mean network response
\begin{align}
    R(\omega;k) = \operatorname{tr}\left(G(\omega;k) G(\omega;k)^H\right),
    \label{eq:energy}
\end{align}
where $\operatorname{tr}(\cdot)$ denotes the matrix trace.
Since~\eqref{eq:energy} is averaged over forcings, it
depends only on the spectrum and not on the eigenmodes,
and can thus be used to tune spectral properties indirectly.  As an instructive example, consider a system with unit mass and only two eigenvalues
at $\omega_{1,2}^2$. The minimum of $R(\omega) = (\omega^2_1 - \omega^2)^{-2} + (\omega^2_2 - \omega^2)^{-2}$ occurs at
$\omega_*^2 = (\omega_1^2 + \omega_2^2)/2$ with the
value $R(\omega_*) = 8(\omega_2^2 - \omega_1^2)^{-2}$,
inversely proportional to the gap width.
Thus, minimizing the response~\eqref{eq:energy} at a
frequency between two eigenvalues will maximize the gap width.
The above framework can be easily adapted to other classes of forcing ensembles, allowing
additional optimization for application-specific input correlations~$\langle \mathbf{F}\mathbf{F}^H\rangle$~\cite{Ronellenfitsch2018a}.

\subsection*{Periodic structures}
The generalization to periodic crystals is straightforward in a Bloch basis, taking the lattice Fourier transform~\cite{Lubensky2015} of the above relations (Supplementary Material). In this case, the trace in~\eqref{eq:energy} is replaced by a sum of traces over the response functions $G_{\mathbf{q}}$ at each wavevector~$\mathbf{q}$ in the first
Brillouin zone of the reciprocal crystal lattice. The Fourier transformed eigenmode equation is then $\hat{K}(\mathbf q) \hat{\mathbf u}_i(\mathbf q) =
m \omega^2_i(\mathbf q) \hat{\mathbf u}_i(\mathbf q)$.
To numerically tune a gap between $\omega_i$ and $\omega_{i+1}$ at wavevector~$\mathbf{q}$, we minimize the objective function
\begin{align}
  \mathcal{L}_i(\mathbf q;k) = R\left[\sqrt{\frac{1}{2}\left(\omega_i^{(0)}(\mathbf{q})^2 +
  \omega_{i+1}^{(0)}(\mathbf{q})^2\right)};k
  \right],
  \label{eq:objective}
\end{align}
where the frequencies $\omega_i^{(0)}$ are evaluated at the initial
stiffness guess $k_e^{(0)}$. Compared with direct gap optimization schemes~\cite{Men2010},
our LRO objective in~\eqref{eq:objective} has the benefit of being differentiable everywhere,
so that it can be efficiently minimized with derivative-based optimization algorithms (Methods).

\begin{figure*}[t]
\centering
\includegraphics[width=\textwidth]{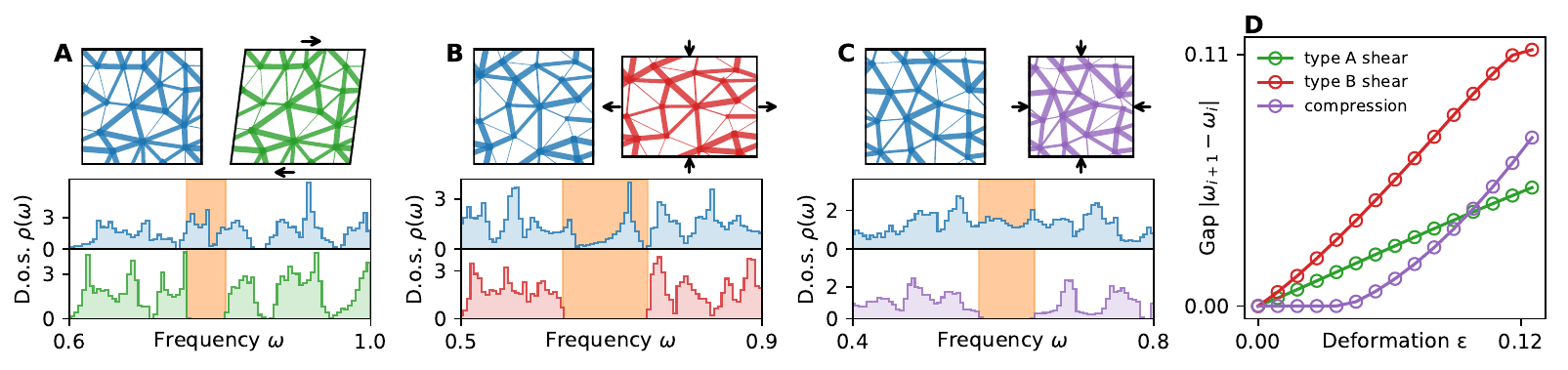}
\caption{\textbf{Phononic metamaterial switches with unit cells possessing random topologies.}
(\textbf{A})~Network designed to open a
bandgap when global type A shear in the $x$ direction
($x \mapsto x + \varepsilon y$, $y\mapsto y$, $\varepsilon=0.125$) is applied (Movie~2).
The gapless unstrained and gapped strained configurations
can be seen in the density of states $\rho(\omega)$.
(\textbf{B})~Network designed to open a gap under global type B
shear ($x\mapsto (1+\varepsilon)x$, $y\mapsto (1-\varepsilon)y$, $\varepsilon=0.125$); see Movie~3.
(\textbf{C})~~Network designed to open a gap  under global compression
($x\mapsto (1-\varepsilon)x$, $y\mapsto (1-\varepsilon)y$, $\varepsilon=0.125$); see Movie~4.
(\textbf{D})~The gap widths $|\omega_{i+1} - \omega_i|$ increase with the
magnitude $\varepsilon$ of the applied deformations in \textbf{A}--\textbf{C}. The linear approximation for the deformed
equilibrium breaks down past $\varepsilon\approx 0.125$.
In all cases, the target deformation was $\varepsilon=0.2$.
Gap sizes were computed by discretizing the entire Brillouin zone using $51 \times 51$ samples.
\label{fg:figure2}}
\end{figure*}
\subsection*{Bandgap tuning of 2D and 3D networks}
The discrete LRO framework is equally applicable to regular and  amorphous network topologies in 2D as well as in 3D. In the examples shown in Fig.~\ref{fg:figure1}\textbf{A-C}, we optimized spring constants $k_e$ over the range $[0.1,1]$ to create one or more bandgaps at predetermined positions in the acoustic spectrum. Sufficiently large networks can support a substantial number of tuned gaps  (Fig.~\ref{fg:figure1}\textbf{D} and Supplementary Material). The characteristics of the final gap-optimized structures do not significantly depend on the base network topology. As general design rule, the distribution of stiffness values $k_e$ in the optimized networks becomes bimodal and peaked at the interval boundaries (Fig.~\ref{fg:figure1}\textbf{A-C}). Intuitively, this means that the LRO algorithm generates impedance mismatches between high- and low-stiffness components.   At higher gap frequencies, the relative proportion of stiff and weak bonds changes such that the mean stiffness decreases independent of the unit cell size  (Fig.~\ref{fg:figure1}\textbf{E}), leading to increasingly more disordered network patterns (Supplementary Material). Thus, low-lying gaps are realized  by
large high-stiffness regions with low-stiffness inclusions whereas high-lying gaps are realized by large low-stiffness regions
with high-stiffness inclusions.

\subsection*{Regular vs. amorphous network topologies}
A practically important question is how many gaps a given network can support.
To explore this question systematically, we attempted to tune up to 32 gaps at random frequencies into more than
$6,000$ networks with different unit cell sizes and base topologies.
Starting from uniform random initial conditions $k_e\in[0.1,1]$, we estimated the success probability $P[\min_{\mathbf{q}}\omega_{i+1}(\mathbf{q}) > \max_{\mathbf{q}}\omega_{i}(\mathbf{q}) \text{ for all gaps } i]$ by explicitly computing the
actual final gap sizes  at $4\times4=16$ different sampling points in the Brillouin zones. As expected, we found
that larger unit cells can support more gaps (Fig.~\ref{fg:figure1}\textbf{D}). Interestingly, however, it is easier to implement a relatively larger number of gaps when the unit cell has randomized vertex positions (Fig.~\ref{fg:figure1}\textbf{D}). Recent work~\cite{Goodrich2015,Rocks2017} showed that amorphous networks are better suited for tuning static elastic properties than regular lattices. Our results indicate that the same is true for the inverse design and control of dynamical properties such as sound transmission.

\subsection*{Robustness}
Notwithstanding the recent major technological advances~\cite{Bhattacharjee2016,Buckmann2012}, fabrication
of discrete metamaterials can be expected to introduce small-to-moderate deviations from the optimal network structure. To demonstrate the robustness of the inversely designed networks, we tuned a single gap at different positions into the frequency spectrum  of networks with different  unit cell topologies.  Thereafter, we perturbed the optimized stiffnesses by adding normally distributed noise (mean 0, standard deviation $\sigma$), and computed the gap size for the perturbed network. Independent  of the specific gap position and network topology,  we found that the gap size decreases as $\sigma$ increases, roughly halving in size as $\sigma$ approaches the lower stiffness bound (Fig.~\ref{fg:figure1}\textbf{F}).  This implies that the low-stiffness components are essential for the realization of both high-lying and low-lying gaps.

\subsection*{Designing phononic switches}
Going beyond basic bandgap tuning, a longstanding unsolved challenge has been the inverse design of metamaterials that adjust their spectra on-demand in response to an external control stimulus. Providing a solution to this problem, we now demonstrate how the above LRO framework can be adapted to design phononic switches that can selectively open and close spectral gaps in pre-programmed frequency ranges (Fig.~\ref{fg:figure2}). As the switching mechanisms we choose global deformations, which have been used previously to induce and control gaps~\cite{Bertoldi2008,Gui2008,Wang2012}. Our approach utilizes the fact that the combination of non-affine network  response and non-zero spring tensions in the strained equilibrium causes systematic changes in the vibrational spectrum. More precisely,
the deformed stiffness matrix $K_\mathrm{def}$ of a spring network under a global deformation $\Gamma: \mathbf{x} \mapsto \Gamma \mathbf{x}$ can be found to lowest order by computing the strained equilibrium positions of all
nodes from the linear dynamics~\eqref{eq:dynamics}, and then expanding the non-linear~\eqref{eq:nonlin} around the strained
equilibrium, removing all linear terms (Methods). From an algorithmic perspective, switch tuning falls into the class of
multi-objective optimization problems which means that, in general, there exists not a single optimal solution but instead a
Pareto front of optimal trade-offs between the individual objectives~\cite{Miettinen1998}.
Here, we parametrize the problem of simultaneously tuning a gap in the deformed state and no gap in the undeformed state
using a no-preference method by considering the scalar least-squares problem,
\begin{align}
    \mathcal{L}_\mathrm{switch}(k) = R_{\mathrm{def}}(\omega^{(0)};k)^2
    + \alpha \left[R(\omega^{(0)};k) - \beta R^{(0)}\right]^2,
    \label{eq:switch}
\end{align}
where $R^{(0)}$ is the response of the
undeformed initial network,
$R_{\mathrm{def}}(\omega^{(0)};k)$ is the response of the deformed network, and $\alpha \in \{0, 1\}$.
The parameter $\beta$ controls the desired response of the undeformed
network. Equation~(\ref{eq:switch}) is minimized in three passes. First,
we set $\alpha=0$, creating a network with a large gap in the
deformed state. Generically, this gap persists when the
deformation is switched off, only decreasing in size.
This persistence is mitigated by running two additional passes
with $\alpha=1$, always feeding in the result
of the previous optimization as initial guess for the next
and recomputing $\omega^{(0)}$ and $R^{(0)}$,
leading to a step-wise increase of the response in the undeformed state
while retaining low response in the deformed network.
The value of $\beta$ such that the network possesses a closed
band structure in the undeformed state and a spectral
gap in the deformed state is found by a parameter search,
and generally lies between $5 \lesssim \beta \lesssim 25$.
Examples of phononic bandgap switches controlled by
the two types of shear transformations and compression  are shown in
Fig.~\ref{fg:figure2} and Supplementary Movies 2--4.

\section*{Discussion}

\subsection*{Continuum elastica}
Macroscopic real-world elastic networks generally exhibit behavior more complex
than pure stretching. Elastic rods can also bend and
twist~\cite{Audoly2010}, and hinge-like connections can significantly
influence the total elastic response. Generalizing the above ideas, we studied
the influence of bending modes and hinges on spectral network design. Through a direct comparison with FE computations,
we found that these two effects suffice to design realistic 3D metamaterial dynamics, as twisting is negligible in the
low frequency regime. Our extended 2D network model treats  bending and hinge stiffness on the same footing by introducing a local preferred relative orientation for elastic rods~\cite{Bergou2008} that are linked by a joint node (Supplementary Material).  Assuming stiff hinges, angular deviation from the preferred orientation
is penalized at each node, and bond bending is modeled by inserting an additional node at
the center of each bond.  As planar continuum realizations of these idealized 2D networks, we performed FE simulations of discrete mechanical metamaterials consisting of
small discs connected by rods of different in-plane diameter (Fig.~\ref{fg:figure3}, Supplementary Material).  The rods are tapered towards their ends to ensure
similar elastic properties near the hinges. The disc-rod-networks are
extruded in the direction normal to their plane to obtain centimeter-thick
quasi-2D material structures (Fig.~\ref{fg:figure3}\textbf{A},\textbf{B}), as can
be 3D printed or cut out of a slab of elastic material. Considering typical Styrodur parameters, we matched the
effective elastic constants of the idealized 2D network model  to those of the 3D FE model (Methods).
\begin{figure}[h]
\centering
\includegraphics[width=\columnwidth]{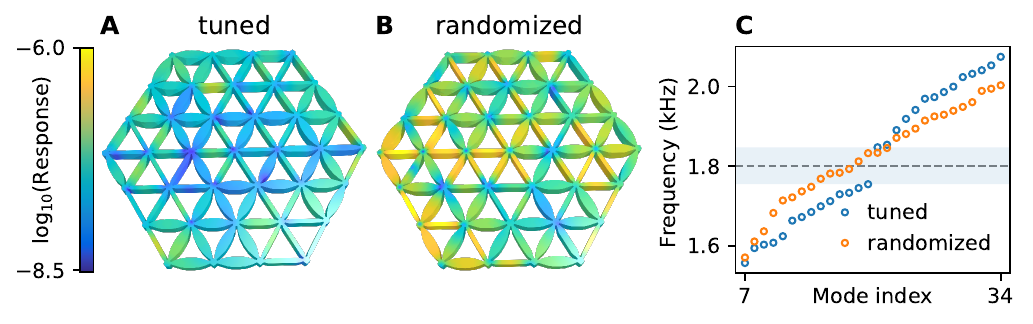}
\caption{\textbf{Spectral gap design for continuum elastic networks.}
(\textbf{A}, \textbf{B}) 3D renderings of two continuum Styrodur networks (diameter $\approx 17\,\mathrm{cm}$, thickness $1\,\mathrm{cm}$) studied in finite element (FE) simulations (Materials and Methods). The color represents the in-plane
harmonic response amplitude $|G_{0j}|$, where the finite elements $j$ respond to harmonic forcing of
the element $0$ at the center of the network with amplitude $(1, 1, 0)/\sqrt{2}$ at the mid-gap frequency $\omega = 1.8\,\mathrm{kHz}$.
The response of the tuned network (\textbf{A}) is significantly smaller than that of the network  (\textbf{B}) with randomly permuted stiffnesses.
(\textbf{C}) FE modes for the tuned network (\textbf{A}) exhibit a substantial spectral gap (blue). The gap vanishes for the network (\textbf{B}) with randomly permuted stiffnesses (orange).
\label{fg:figure3}}
\end{figure}
The reduction to the 2D model for band gap design is made possible by the fact that, for sufficiently thick 3D materials,
the spectrum is divided into in-plane and out-of-plane modes. Since in-plane and out-of-plane dynamics are approximately decoupled in this regime, it suffices to optimize the spectral gaps associated with these in-plane modes (Supplementary Material).  By tuning a low-frequency spectral gap  into the 2D network and mapping back onto the full 3D FE model (Fig.~\ref{fg:figure3}\textbf{A}),
we find that the gap remains highly conserved in the FE mode spectrum  (Fig.~\ref{fg:figure3}\textbf{C}, blue circles). In contrast,  a control network (Fig.~\ref{fg:figure3}\textbf{B}) obtained by randomly permuting the edges of the optimized network  loses the gap (Fig.~\ref{fg:figure3}\textbf{C}, orange circles). As a consequence, the gapped 3D continuum network exhibits a significantly reduced response when the center node is forced at mid-gap frequency (Fig.~\ref{fg:figure3}\textbf{A},\textbf{B}). This demonstration illustrates the vast practical potential of the LRO approach with regard to the inverse design of complex 3D metamaterial functionalities.

\subsection*{Topologically protected modes}
\begin{figure*}[t]
    \centering
    \includegraphics[width=\textwidth]{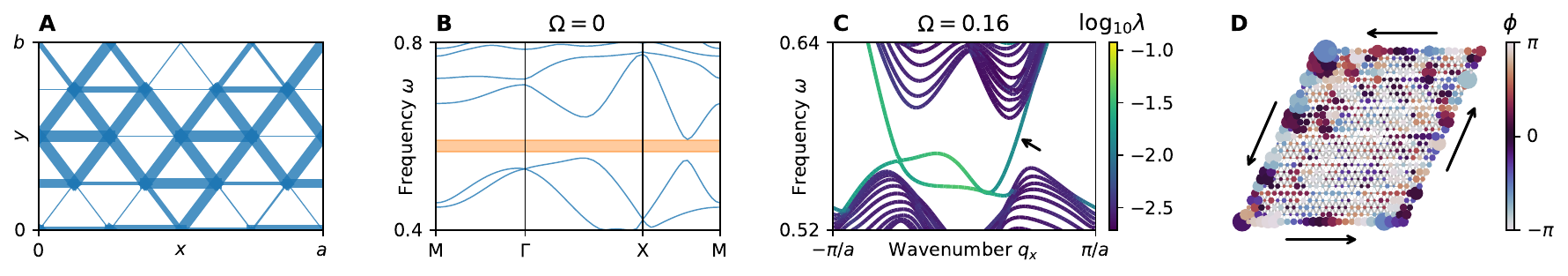}
    \caption{Protected chiral edge modes in designed networks.
    (\textbf{A})~Designed gapped network  on a
    triangular lattice.
    (\textbf{B})~Gapped band structure of the
    network from \textbf{A} in the symmetry-unbroken
    state ($\Omega = 0$, Chern invariant $\Delta C=0$).
    (\textbf{C})~Chiral protected edge band (arrow) in the topological phase ($\Omega=0.16$,
    Chern invariant $\Delta C = 1$)
    for a ribbon with 12 open-boundary unit cells in the $y$ direction
    and periodic boundary conditions in the $x$ direction.
    Color corresponds to the mode localization ratio
    $\lambda = \sum_i u_i^4/(\sum_i u_i^2)^2$.
    The topological phase transition occurs at
    $\Omega_c \approx 0.08$.
    (\textbf{D})~Localized chiral edge response to forcing
    in the lower left corner in a finite sample network
    with $6\times 6$ unit cells. Circle size corresponds
    to the norm $\|\mathbf{u}_i\|^2$ of the linear response
    at long times, color corresponds to the response phase.}
    \label{fg:figure4}
\end{figure*}
Topological mechanics offers a powerful framework for the
control of phononic excitations. Topologically protected
phonons localized at the edges of gapped mechanical materials
have been studied extensively in special lattices~\cite{Kane2014,Nash2015,Souslov2016,Wang2015,Mitchell2018} and in random networks~\cite{Agarwala2017,Mitchell2018b}. Here,
 we demonstrate that our designed networks can harbor such
modes as well. Inverse design thus promises precise control of topological materials
and may pave the way to fully programmable topology.
Specifically, we consider 2D Chern insulators,
originally discovered through the Quantum Hall effect, in which
protected modes arise through the breaking of time-reversal
invariance. A mechanical Chern material was realized recently with active gyroscopes~\cite{Wang2015,Nash2015}. We focus on the direct
analog of Hall insulators, breaking time-reversal symmetry by rotating a gap-tuned network about an axis perpendicular to
it~\cite{Wang2015c}.
The resulting Coriolis force plays the role of an external magnetic
field. To linear order in the rotation frequency $\Omega$, the equations of
motion read
\begin{align}
m\,\ddot{\mathbf{u}} + 2m\,\mathbf{\Omega} \times \dot{\mathbf{u}} + K \mathbf{u} = 0,
\end{align}
where $\mathbf{\Omega} \times \mathbf{x} = (-\Omega\, x_2, \Omega\, x_1)$ in the 2D plane. By increasing the rotation rate $\Omega$, a 2D network with designed bandgap (Fig.~\ref{fg:figure4}\textbf{A})
can be driven through a topological phase transition. The initially open
gap (Fig.~\ref{fg:figure4}\textbf{B}) closes at some finite
$\Omega = \Omega_c$, and then re-opens
in a topological phase with non-zero bulk Chern invariant,
revealing a localized band of protected edge modes in the gap
(Fig.~\ref{fg:figure4}\textbf{C}).
In finite samples, the non-trivial topology leads to the appearance
of robust, protected edge modes allowing unidirectional transport of
phonons along the sample boundary (Fig.~\ref{fg:figure4}\textbf{D}),
with chirality controlled by the
sign of $\Omega$. Although not every tuned gap can be made topological in this manner,
our results suggest that topological mechanical materials could
be programmed to exhibit protected modes at almost any frequency, gap size,
and chirality by tuning the basic building blocks of the
underlying network.

\section*{Conclusions}

We showed that linear response optimization (LRO) provides a flexible and efficient algorithmic
framework for the inverse design of discrete elastic metamaterials with desirable spectral properties.
Notably, LRO opens a path towards developing and exploring materials with amorphous unit cells, which
appear capable of hosting a significantly larger number of programmable frequency gaps than their more commonly studied
highly symmetric counterparts (Fig.~\ref{fg:figure1}).  Furthermore, LRO makes it possible to rationally design mechanical
systems with switchable band structure that can be controlled through external deformations (Fig.~\ref{fg:figure2}). Disordered base structures are particularly promising candidates in this context, because their non-affine responses facilitate large spectral differences between deformed and undeformed material configurations. Another intriguing prospect is the inverse design of topological materials with highly controllable mode protection (Fig.~\ref{fg:figure4}). 
\par
The LRO approach developed here can be easily adapted to impose  desired bandgaps at different selected positions in the Brillouin zone, or to realize more complex spectral features such as degenerate points.  While the present study focused on optimizing the global network response via~\eqref{eq:energy}, we anticipate that the tuning of individual covariance matrix elements can provide a powerful technique for  the fine-grained control of specific dynamical properties.  Perhaps most importantly, however, due to its generic mathematical formulation, the LRO scheme can be applied  to any dynamical system that can be linearized near a stable fixed point.
This promises exciting future  possibilities for the inverse design of complex static and dynamic target functionalities in a wide range of systems,  from amorphous photonic crystals~\cite{Ricouvier2019} and DNA origami structures~\cite{Bathe2017,Sun2014} to passive~\cite{Lee2017} and active~\cite{Kotwal2019} topolectrical circuits.

\section*{Methods}
\subsection*{Gradient-based optimization}
We employed the standard L-BFGS-B bound-constrained quasi-Newton
algorithm~\cite{Byrd1995}. In the numerical examples,
stiffness values $k_e$ were optimized over the interval $[0.1, 1]$.

\subsection*{Randomized topology} 
Networks with randomized topology were constructed
by first distributing points $\mathbf x_i$ in the periodic unit cell uniformly at random,
and then locally minimizing the potential function
$V = \sum_{i\neq j}d(\mathbf x_i, \mathbf x_j)^{-1}$ where
the distance $d(\mathbf x_i, \mathbf x_j)^{-1} = \min_{\mathbf m}
\| \mathbf x_i - \mathbf x_j + \mathbf m \|$ for lattice vectors
$\mathbf m$.
The basic unit cell points were copied into the directly adjacent unit cells and
a Delaunay triangulation was calculated to obtain the network topology
and unit vectors along the bonds.
Equivalent points were identified in the basic unit cell
and the adjacent unit cells deleted.

\subsection*{Spectra of deformed networks} 
The spectrum of an
elastic network described by \eqref{eq:nonlin} is found by linearizing
the elastic energy to
\begin{align}
    V = \frac{1}{2} \sum_e k_e \left( \hat{{\mathbf b}}_e \cdot \Delta \mathbf{u}_e\right)^2,
    \label{eq:linearized}
\end{align}
where $k_e$ are the spring stiffnesses of spring $e$, $\hat{{\mathbf b}}_e$ is
the unit vector pointing along $e=(ij)$, and $\Delta \mathbf{u}_e =
\mathbf{u}_i - \mathbf{u}_j$ for the small displacements
$\mathbf{x}_i = \mathbf{x}_i^{(0)} + \mathbf{u}_i$ from the equilibrium
positions $\mathbf{x}_i^{(0)}$.
To linear order, a global deformation is then defined by
\begin{align}
  \mathbf{x} &\rightarrow (\mathbb{1} + \eta)(\mathbf{x}^{(0)} + \mathbf{u}) \approx \mathbf{x}^{(0)} + \eta \mathbf{x}^{(0)} + \mathbf{u},
\end{align}
where the $d\times d$ matrix $\eta$ encodes the global deformation.
Plugging this into \eqref{eq:linearized} and minimizing with respect
to $\mathbf{u}$, we obtain the new equilibrium positions of the network
nodes under the global deformation.

The spectrum of the deformed network is then found by expanding \eqref{eq:linearized} around the new equilibrium positions
$\mathbf{x}_\mathrm{def} = \mathbf{x}^{(0)} + \eta \mathbf{x}^{(0)} + \mathbf{u}_\mathrm{def}$.
Up to a constant we obtain
\begin{align}
  V_\mathrm{def} = \frac{1}{2} \sum_{e} k_e \left[
  \frac{\ell_e^{(0)}}{\tilde\ell_e} \left(\tilde{\mathbf{b}}_e \cdot \Delta\mathbf{v}_e\right)^2
  + \left(1 - \frac{\ell_e^{(0)}}{\tilde\ell_e} \right) (\Delta\mathbf{v}_e)^2
  \right],
\end{align}
where the tilde denotes a quantity in the deformed equilibrium,
and $\mathbf{v}_i$ are the small displacements around the deformed equilibrium.

\subsection*{Finite Element (FE) calculations} 
We used \textsc{MATLAB} 2018b's
\texttt{createpde} command in the \texttt{structural} mode for
solid modal analysis. The meshes were generated using the default
parameters, only the maximum mesh size was set to \texttt{0.005}.
The material properties were set for Styrodur (BASF AG, Ludwigshafen, Germany),
a polymer foam
with Young's modulus $Y=2\cdot 10^7 \,\mathrm{Pa}$, Poisson's ratio $\nu=0.4$, and
density $\rho = 33 \,\mathrm{kg}/\mathrm{m}^3$.
To compare the results of 3D FE calculations to
our 2D network model, for each mode
with $x$,$y$,$z$ components $(\mathbf{u}_x, \mathbf{u}_y,
\mathbf{u}_z)$  we computed the in-plane
contribution $f_{xy}^2 = (\|\mathbf{u}_x\|^2 + \|\mathbf{u}_y\|^2)/
(\|\mathbf{u}_x\|^2 + \|\mathbf{u}_y\|^2 + \|\mathbf{u}_z\|^2)$,
and discarded all modes with $f_{xy} < 0.5$. Additional details of the mapping between 
the 2D network model  and 3D FE model are provided in the Supplementary Material.


\bibliography{bandgaps}

\begin{thebibliography}{51}%
\makeatletter
\providecommand \@ifxundefined [1]{%
 \@ifx{#1\undefined}
}%
\providecommand \@ifnum [1]{%
 \ifnum #1\expandafter \@firstoftwo
 \else \expandafter \@secondoftwo
 \fi
}%
\providecommand \@ifx [1]{%
 \ifx #1\expandafter \@firstoftwo
 \else \expandafter \@secondoftwo
 \fi
}%
\providecommand \natexlab [1]{#1}%
\providecommand \enquote  [1]{``#1''}%
\providecommand \bibnamefont  [1]{#1}%
\providecommand \bibfnamefont [1]{#1}%
\providecommand \citenamefont [1]{#1}%
\providecommand \href@noop [0]{\@secondoftwo}%
\providecommand \href [0]{\begingroup \@sanitize@url \@href}%
\providecommand \@href[1]{\@@startlink{#1}\@@href}%
\providecommand \@@href[1]{\endgroup#1\@@endlink}%
\providecommand \@sanitize@url [0]{\catcode `\\12\catcode `\$12\catcode
  `\&12\catcode `\#12\catcode `\^12\catcode `\_12\catcode `\%12\relax}%
\providecommand \@@startlink[1]{}%
\providecommand \@@endlink[0]{}%
\providecommand \url  [0]{\begingroup\@sanitize@url \@url }%
\providecommand \@url [1]{\endgroup\@href {#1}{\urlprefix }}%
\providecommand \urlprefix  [0]{URL }%
\providecommand \Eprint [0]{\href }%
\providecommand \doibase [0]{https://doi.org/}%
\providecommand \selectlanguage [0]{\@gobble}%
\providecommand \bibinfo  [0]{\@secondoftwo}%
\providecommand \bibfield  [0]{\@secondoftwo}%
\providecommand \translation [1]{[#1]}%
\providecommand \BibitemOpen [0]{}%
\providecommand \bibitemStop [0]{}%
\providecommand \bibitemNoStop [0]{.\EOS\space}%
\providecommand \EOS [0]{\spacefactor3000\relax}%
\providecommand \BibitemShut  [1]{\csname bibitem#1\endcsname}%
\let\auto@bib@innerbib\@empty
\bibitem [{\citenamefont {Setyawan}\ and\ \citenamefont
  {Curtarolo}(2010)}]{Setyawan2010}%
  \BibitemOpen
  \bibfield  {author} {\bibinfo {author} {\bibfnamefont {W.}~\bibnamefont
  {Setyawan}}and\ \bibinfo {author} {\bibfnamefont {S.}~\bibnamefont
  {Curtarolo}},\ }\bibfield  {title} {\bibinfo {title} {{High-throughput
  electronic band structure calculations: Challenges and tools}},\ }\href
  {https://doi.org/10.1016/j.commatsci.2010.05.010} {\bibfield  {journal}
  {\bibinfo  {journal} {Comput. Mater. Sci.}\ }\textbf {\bibinfo {volume}
  {49}},\ \bibinfo {pages} {299} (\bibinfo {year} {2010})},\ \Eprint
  {https://arxiv.org/abs/1004.2974} {arXiv:1004.2974} \BibitemShut {NoStop}%
\bibitem [{\citenamefont {Cummer}\ \emph {et~al.}(2016)\citenamefont {Cummer},
  \citenamefont {Christensen},\ and\ \citenamefont {Al{\`{u}}}}]{Cummer2016}%
  \BibitemOpen
  \bibfield  {author} {\bibinfo {author} {\bibfnamefont {S.~A.}\ \bibnamefont
  {Cummer}}, \bibinfo {author} {\bibfnamefont {J.}~\bibnamefont {Christensen}},
  and\ \bibinfo {author} {\bibfnamefont {A.}~\bibnamefont {Al{\`{u}}}},\
  }\bibfield  {title} {\bibinfo {title} {{Controlling sound with acoustic
  metamaterials}},\ }\href@noop {} {\bibfield  {journal} {\bibinfo  {journal}
  {Nat. Rev. Mater.}\ }\textbf {\bibinfo {volume} {1}},\ \bibinfo {pages}
  {16001} (\bibinfo {year} {2016})}\BibitemShut {NoStop}%
\bibitem [{\citenamefont {Zhang}\ \emph {et~al.}(2011)\citenamefont {Zhang},
  \citenamefont {Xia},\ and\ \citenamefont {Fang}}]{Zhang2011}%
  \BibitemOpen
  \bibfield  {author} {\bibinfo {author} {\bibfnamefont {S.}~\bibnamefont
  {Zhang}}, \bibinfo {author} {\bibfnamefont {C.}~\bibnamefont {Xia}}, and\
  \bibinfo {author} {\bibfnamefont {N.}~\bibnamefont {Fang}},\ }\bibfield
  {title} {\bibinfo {title} {{Broadband Acoustic Cloak for Ultrasound Waves}},\
  }\href {https://doi.org/10.1103/PhysRevLett.106.024301} {\bibfield  {journal}
  {\bibinfo  {journal} {Phys. Rev. Lett.}\ }\textbf {\bibinfo {volume} {106}},\
  \bibinfo {pages} {024301} (\bibinfo {year} {2011})},\ \Eprint
  {https://arxiv.org/abs/1009.3310} {arXiv:1009.3310} \BibitemShut {NoStop}%
\bibitem [{\citenamefont {Brunet}\ \emph {et~al.}(2014)\citenamefont {Brunet},
  \citenamefont {Merlin}, \citenamefont {Mascaro}, \citenamefont {Zimny},
  \citenamefont {Leng}, \citenamefont {Poncelet}, \citenamefont
  {Arist{\'{e}}gui},\ and\ \citenamefont {Mondain-Monval}}]{Brunet2014}%
  \BibitemOpen
  \bibfield  {author} {\bibinfo {author} {\bibfnamefont {T.}~\bibnamefont
  {Brunet}}, \bibinfo {author} {\bibfnamefont {A.}~\bibnamefont {Merlin}},
  \bibinfo {author} {\bibfnamefont {B.}~\bibnamefont {Mascaro}}, \bibinfo
  {author} {\bibfnamefont {K.}~\bibnamefont {Zimny}}, \bibinfo {author}
  {\bibfnamefont {J.}~\bibnamefont {Leng}}, \bibinfo {author} {\bibfnamefont
  {O.}~\bibnamefont {Poncelet}}, \bibinfo {author} {\bibfnamefont
  {C.}~\bibnamefont {Arist{\'{e}}gui}}, and\ \bibinfo {author} {\bibfnamefont
  {O.}~\bibnamefont {Mondain-Monval}},\ }\bibfield  {title} {\bibinfo {title}
  {{Soft 3D acoustic metamaterial with negative index}},\ }\href
  {https://doi.org/10.1038/nmat4164} {\bibfield  {journal} {\bibinfo  {journal}
  {Nat. Mater.}\ }\textbf {\bibinfo {volume} {14}},\ \bibinfo {pages} {384}
  (\bibinfo {year} {2014})}\BibitemShut {NoStop}%
\bibitem [{\citenamefont {Mart{\'{i}}nez-Sala}\ \emph
  {et~al.}(1995)\citenamefont {Mart{\'{i}}nez-Sala}, \citenamefont {Sancho},
  \citenamefont {S{\'{a}}nchez}, \citenamefont {G{\'{o}}mez}, \citenamefont
  {Llinares},\ and\ \citenamefont {Meseguer}}]{Martinez-Sala1995}%
  \BibitemOpen
  \bibfield  {author} {\bibinfo {author} {\bibfnamefont {R.}~\bibnamefont
  {Mart{\'{i}}nez-Sala}}, \bibinfo {author} {\bibfnamefont {J.}~\bibnamefont
  {Sancho}}, \bibinfo {author} {\bibfnamefont {J.~V.}\ \bibnamefont
  {S{\'{a}}nchez}}, \bibinfo {author} {\bibfnamefont {V.}~\bibnamefont
  {G{\'{o}}mez}}, \bibinfo {author} {\bibfnamefont {J.}~\bibnamefont
  {Llinares}}, and\ \bibinfo {author} {\bibfnamefont {F.}~\bibnamefont
  {Meseguer}},\ }\bibfield  {title} {\bibinfo {title} {{Sound attenuation by
  sculpture}},\ }\href {https://doi.org/10.1038/378241a0} {\bibfield  {journal}
  {\bibinfo  {journal} {Nature}\ }\textbf {\bibinfo {volume} {378}},\ \bibinfo
  {pages} {241} (\bibinfo {year} {1995})}\BibitemShut {NoStop}%
\bibitem [{\citenamefont {Bhattacharjee}\ \emph {et~al.}(2016)\citenamefont
  {Bhattacharjee}, \citenamefont {Urrios}, \citenamefont {Kang},\ and\
  \citenamefont {Folch}}]{Bhattacharjee2016}%
  \BibitemOpen
  \bibfield  {author} {\bibinfo {author} {\bibfnamefont {N.}~\bibnamefont
  {Bhattacharjee}}, \bibinfo {author} {\bibfnamefont {A.}~\bibnamefont
  {Urrios}}, \bibinfo {author} {\bibfnamefont {S.}~\bibnamefont {Kang}}, and\
  \bibinfo {author} {\bibfnamefont {A.}~\bibnamefont {Folch}},\ }\bibfield
  {title} {\bibinfo {title} {The upcoming 3d-printing revolution in
  microfluidics},\ }\href@noop {} {\bibfield  {journal} {\bibinfo  {journal}
  {Lab Chip}\ }\textbf {\bibinfo {volume} {16}},\ \bibinfo {pages} {1720}
  (\bibinfo {year} {2016})}\BibitemShut {NoStop}%
\bibitem [{\citenamefont {B{\"{u}}ckmann}\ \emph {et~al.}(2012)\citenamefont
  {B{\"{u}}ckmann}, \citenamefont {Stenger}, \citenamefont {Kadic},
  \citenamefont {Kaschke}, \citenamefont {Fr{\"{o}}lich}, \citenamefont
  {Kennerknecht}, \citenamefont {Eberl}, \citenamefont {Thiel},\ and\
  \citenamefont {Wegener}}]{Buckmann2012}%
  \BibitemOpen
  \bibfield  {author} {\bibinfo {author} {\bibfnamefont {T.}~\bibnamefont
  {B{\"{u}}ckmann}}, \bibinfo {author} {\bibfnamefont {N.}~\bibnamefont
  {Stenger}}, \bibinfo {author} {\bibfnamefont {M.}~\bibnamefont {Kadic}},
  \bibinfo {author} {\bibfnamefont {J.}~\bibnamefont {Kaschke}}, \bibinfo
  {author} {\bibfnamefont {A.}~\bibnamefont {Fr{\"{o}}lich}}, \bibinfo {author}
  {\bibfnamefont {T.}~\bibnamefont {Kennerknecht}}, \bibinfo {author}
  {\bibfnamefont {C.}~\bibnamefont {Eberl}}, \bibinfo {author} {\bibfnamefont
  {M.}~\bibnamefont {Thiel}}, and\ \bibinfo {author} {\bibfnamefont
  {M.}~\bibnamefont {Wegener}},\ }\bibfield  {title} {\bibinfo {title}
  {{Tailored 3D mechanical metamaterials made by dip-in direct-laser-writing
  optical lithography}},\ }\href@noop {} {\bibfield  {journal} {\bibinfo
  {journal} {Adv. Mater.}\ }\textbf {\bibinfo {volume} {24}},\ \bibinfo {pages}
  {2710} (\bibinfo {year} {2012})}\BibitemShut {NoStop}%
\bibitem [{\citenamefont {Bertoldi}\ \emph {et~al.}(2017)\citenamefont
  {Bertoldi}, \citenamefont {Vitelli}, \citenamefont {Christensen},\ and\
  \citenamefont {van Hecke}}]{Bertoldi2017}%
  \BibitemOpen
  \bibfield  {author} {\bibinfo {author} {\bibfnamefont {K.}~\bibnamefont
  {Bertoldi}}, \bibinfo {author} {\bibfnamefont {V.}~\bibnamefont {Vitelli}},
  \bibinfo {author} {\bibfnamefont {J.}~\bibnamefont {Christensen}}, and\
  \bibinfo {author} {\bibfnamefont {M.}~\bibnamefont {van Hecke}},\ }\bibfield
  {title} {\bibinfo {title} {{Flexible mechanical metamaterials}},\ }\href
  {https://doi.org/10.1038/natrevmats.2017.66} {\bibfield  {journal} {\bibinfo
  {journal} {Nat. Rev. Mater.}\ }\textbf {\bibinfo {volume} {2}},\ \bibinfo
  {pages} {17066} (\bibinfo {year} {2017})}\BibitemShut {NoStop}%
\bibitem [{\citenamefont {Ma}\ \emph {et~al.}(2016)\citenamefont {Ma},
  \citenamefont {Fu}, \citenamefont {Wang}, \citenamefont {del Hougne},
  \citenamefont {Christensen}, \citenamefont {Lai},\ and\ \citenamefont
  {Sheng}}]{Ma2016}%
  \BibitemOpen
  \bibfield  {author} {\bibinfo {author} {\bibfnamefont {G.}~\bibnamefont
  {Ma}}, \bibinfo {author} {\bibfnamefont {C.}~\bibnamefont {Fu}}, \bibinfo
  {author} {\bibfnamefont {G.}~\bibnamefont {Wang}}, \bibinfo {author}
  {\bibfnamefont {P.}~\bibnamefont {del Hougne}}, \bibinfo {author}
  {\bibfnamefont {J.}~\bibnamefont {Christensen}}, \bibinfo {author}
  {\bibfnamefont {Y.}~\bibnamefont {Lai}}, and\ \bibinfo {author}
  {\bibfnamefont {P.}~\bibnamefont {Sheng}},\ }\bibfield  {title} {\bibinfo
  {title} {{Polarization bandgaps and fluid-like elasticity in fully solid
  elastic metamaterials}},\ }\href {https://doi.org/10.1038/ncomms13536}
  {\bibfield  {journal} {\bibinfo  {journal} {Nat. Commun.}\ }\textbf {\bibinfo
  {volume} {7}},\ \bibinfo {pages} {13536} (\bibinfo {year}
  {2016})}\BibitemShut {NoStop}%
\bibitem [{\citenamefont {Deymier}(2013)}]{Deymier2013}%
  \BibitemOpen
  \bibfield  {author} {\bibinfo {author} {\bibfnamefont {P.~A.}\ \bibnamefont
  {Deymier}},\ }\href {https://doi.org/10.1007/978-3-642-31232-8} {\emph
  {\bibinfo {title} {Springer series in solid-state sciences}}},\ \bibinfo
  {series} {Springer Series in Solid-State Sciences}, Vol.\ \bibinfo {volume}
  {173}\ (\bibinfo  {publisher} {Springer Berlin Heidelberg},\ \bibinfo
  {address} {Berlin, Heidelberg},\ \bibinfo {year} {2013})\ pp.\ \bibinfo
  {pages} {xiv, 378 p.},\ \Eprint {https://arxiv.org/abs/1206.5023}
  {arXiv:1206.5023} \BibitemShut {NoStop}%
\bibitem [{\citenamefont {S{\"{u}}sstrunk}\ and\ \citenamefont
  {Huber}(2016)}]{Susstrunk2016}%
  \BibitemOpen
  \bibfield  {author} {\bibinfo {author} {\bibfnamefont {R.}~\bibnamefont
  {S{\"{u}}sstrunk}}and\ \bibinfo {author} {\bibfnamefont {S.~D.}\ \bibnamefont
  {Huber}},\ }\bibfield  {title} {\bibinfo {title} {{Classification of
  topological phonons in linear mechanical metamaterials}},\ }\href
  {https://doi.org/10.1073/pnas.1605462113} {\bibfield  {journal} {\bibinfo
  {journal} {Proc. Natl. Acad. Sci. U.S.A.}\ }\textbf {\bibinfo {volume}
  {113}},\ \bibinfo {pages} {E4767} (\bibinfo {year} {2016})},\ \Eprint
  {https://arxiv.org/abs/1604.01033} {arXiv:1604.01033} \BibitemShut {NoStop}%
\bibitem [{\citenamefont {Sigmund}\ and\ \citenamefont {{Sondergaard
  Jensen}}(2003)}]{Sigmund2003}%
  \BibitemOpen
  \bibfield  {author} {\bibinfo {author} {\bibfnamefont {O.}~\bibnamefont
  {Sigmund}}and\ \bibinfo {author} {\bibfnamefont {J.}~\bibnamefont
  {{Sondergaard Jensen}}},\ }\bibfield  {title} {\bibinfo {title} {{Systematic
  design of phononic band-gap materials and structures by topology
  optimization}},\ }\href {https://doi.org/10.1098/rsta.2003.1177} {\bibfield
  {journal} {\bibinfo  {journal} {Philos. Trans. Royal Soc. A}\ }\textbf
  {\bibinfo {volume} {361}},\ \bibinfo {pages} {1001} (\bibinfo {year}
  {2003})}\BibitemShut {NoStop}%
\bibitem [{\citenamefont {Bends{\o}e}\ and\ \citenamefont
  {Sigmund}(2003)}]{Bendsoe2003}%
  \BibitemOpen
  \bibfield  {author} {\bibinfo {author} {\bibfnamefont {M.~P.}\ \bibnamefont
  {Bends{\o}e}}and\ \bibinfo {author} {\bibfnamefont {O.}~\bibnamefont
  {Sigmund}},\ }\bibfield  {title} {\bibinfo {title} {{Topology optimization:
  theory, methods, and applications}},\ }\href
  {http://www.amazon.fr/Topology-Optimization-Theory-Methods-Applications/dp/3540429921}
  {\bibfield  {journal} {\bibinfo  {journal} {Engineering}\ }\textbf {\bibinfo
  {volume} {2nd Edition}},\ \bibinfo {pages} {370} (\bibinfo {year}
  {2003})}\BibitemShut {NoStop}%
\bibitem [{\citenamefont {Diest}(2013)}]{Diest2013}%
  \BibitemOpen
  \bibfield  {author} {\bibinfo {author} {\bibfnamefont {K.}~\bibnamefont
  {Diest}},\ }\href {https://doi.org/10.1007/978-94-007-6664-8} {\emph
  {\bibinfo {title} {{Numerical Methods for Metamaterial Design (Topics in
  Applied Physics)}}}}\ (\bibinfo  {publisher} {Springer},\ \bibinfo {year}
  {2013})\ p.\ \bibinfo {pages} {250}\BibitemShut {NoStop}%
\bibitem [{\citenamefont {Rechtsman}\ \emph {et~al.}(2008)\citenamefont
  {Rechtsman}, \citenamefont {Jeong}, \citenamefont {Chaikin}, \citenamefont
  {Torquato},\ and\ \citenamefont {Steinhardt}}]{Rechtsman2008}%
  \BibitemOpen
  \bibfield  {author} {\bibinfo {author} {\bibfnamefont {M.~C.}\ \bibnamefont
  {Rechtsman}}, \bibinfo {author} {\bibfnamefont {H.-C.}\ \bibnamefont
  {Jeong}}, \bibinfo {author} {\bibfnamefont {P.~M.}\ \bibnamefont {Chaikin}},
  \bibinfo {author} {\bibfnamefont {S.}~\bibnamefont {Torquato}}, and\ \bibinfo
  {author} {\bibfnamefont {P.~J.}\ \bibnamefont {Steinhardt}},\ }\bibfield
  {title} {\bibinfo {title} {{Optimized Structures for Photonic
  Quasicrystals}},\ }\href {https://doi.org/10.1103/PhysRevLett.101.073902}
  {\bibfield  {journal} {\bibinfo  {journal} {Phys. Rev. Lett.}\ }\textbf
  {\bibinfo {volume} {101}},\ \bibinfo {pages} {073902} (\bibinfo {year}
  {2008})}\BibitemShut {NoStop}%
\bibitem [{\citenamefont {Man}\ \emph {et~al.}(2013)\citenamefont {Man},
  \citenamefont {Florescu}, \citenamefont {Matsuyama}, \citenamefont {Yadak},
  \citenamefont {Nahal}, \citenamefont {Hashemizad}, \citenamefont
  {Williamson}, \citenamefont {Steinhardt}, \citenamefont {Torquato},\ and\
  \citenamefont {Chaikin}}]{Man2013}%
  \BibitemOpen
  \bibfield  {author} {\bibinfo {author} {\bibfnamefont {W.}~\bibnamefont
  {Man}}, \bibinfo {author} {\bibfnamefont {M.}~\bibnamefont {Florescu}},
  \bibinfo {author} {\bibfnamefont {K.}~\bibnamefont {Matsuyama}}, \bibinfo
  {author} {\bibfnamefont {P.}~\bibnamefont {Yadak}}, \bibinfo {author}
  {\bibfnamefont {G.}~\bibnamefont {Nahal}}, \bibinfo {author} {\bibfnamefont
  {S.}~\bibnamefont {Hashemizad}}, \bibinfo {author} {\bibfnamefont
  {E.}~\bibnamefont {Williamson}}, \bibinfo {author} {\bibfnamefont
  {P.}~\bibnamefont {Steinhardt}}, \bibinfo {author} {\bibfnamefont
  {S.}~\bibnamefont {Torquato}}, and\ \bibinfo {author} {\bibfnamefont
  {P.}~\bibnamefont {Chaikin}},\ }\bibfield  {title} {\bibinfo {title}
  {{Photonic band gap in isotropic hyperuniform disordered solids with low
  dielectric contrast}},\ }\href@noop {} {\bibfield  {journal} {\bibinfo
  {journal} {Opt. Express}\ }\textbf {\bibinfo {volume} {21}},\ \bibinfo
  {pages} {19972} (\bibinfo {year} {2013})}\BibitemShut {NoStop}%
\bibitem [{\citenamefont {Florescu}\ \emph {et~al.}(2009)\citenamefont
  {Florescu}, \citenamefont {Torquato},\ and\ \citenamefont
  {Steinhardt}}]{Florescu2009}%
  \BibitemOpen
  \bibfield  {author} {\bibinfo {author} {\bibfnamefont {M.}~\bibnamefont
  {Florescu}}, \bibinfo {author} {\bibfnamefont {S.}~\bibnamefont {Torquato}},
  and\ \bibinfo {author} {\bibfnamefont {P.~J.}\ \bibnamefont {Steinhardt}},\
  }\bibfield  {title} {\bibinfo {title} {{Designer disordered materials with
  large, complete photonic band gaps}},\ }\href
  {https://doi.org/10.1073/pnas.0907744106} {\bibfield  {journal} {\bibinfo
  {journal} {Proc. Natl. Acad. Sci. U.S.A.}\ }\textbf {\bibinfo {volume}
  {106}},\ \bibinfo {pages} {20658} (\bibinfo {year} {2009})},\ \Eprint
  {https://arxiv.org/abs/1007.3554} {arXiv:1007.3554} \BibitemShut {NoStop}%
\bibitem [{\citenamefont {Mousanezhad}\ \emph {et~al.}(2015)\citenamefont
  {Mousanezhad}, \citenamefont {Babaee}, \citenamefont {Ghosh}, \citenamefont
  {Mahdi}, \citenamefont {Bertoldi},\ and\ \citenamefont
  {Vaziri}}]{Mousanezhad2015}%
  \BibitemOpen
  \bibfield  {author} {\bibinfo {author} {\bibfnamefont {D.}~\bibnamefont
  {Mousanezhad}}, \bibinfo {author} {\bibfnamefont {S.}~\bibnamefont {Babaee}},
  \bibinfo {author} {\bibfnamefont {R.}~\bibnamefont {Ghosh}}, \bibinfo
  {author} {\bibfnamefont {E.}~\bibnamefont {Mahdi}}, \bibinfo {author}
  {\bibfnamefont {K.}~\bibnamefont {Bertoldi}}, and\ \bibinfo {author}
  {\bibfnamefont {A.}~\bibnamefont {Vaziri}},\ }\bibfield  {title} {\bibinfo
  {title} {{Honeycomb phononic crystals with self-similar hierarchy}},\ }\href
  {https://doi.org/10.1103/PhysRevB.92.104304} {\bibfield  {journal} {\bibinfo
  {journal} {Phys. Rev. B}\ }\textbf {\bibinfo {volume} {92}},\ \bibinfo
  {pages} {104304} (\bibinfo {year} {2015})}\BibitemShut {NoStop}%
\bibitem [{\citenamefont {Nash}\ \emph {et~al.}(2015)\citenamefont {Nash},
  \citenamefont {Kleckner}, \citenamefont {Read}, \citenamefont {Vitelli},
  \citenamefont {Turner},\ and\ \citenamefont {Irvine}}]{Nash2015}%
  \BibitemOpen
  \bibfield  {author} {\bibinfo {author} {\bibfnamefont {L.~M.}\ \bibnamefont
  {Nash}}, \bibinfo {author} {\bibfnamefont {D.}~\bibnamefont {Kleckner}},
  \bibinfo {author} {\bibfnamefont {A.}~\bibnamefont {Read}}, \bibinfo {author}
  {\bibfnamefont {V.}~\bibnamefont {Vitelli}}, \bibinfo {author} {\bibfnamefont
  {A.~M.}\ \bibnamefont {Turner}}, and\ \bibinfo {author} {\bibfnamefont
  {W.~T.~M.}\ \bibnamefont {Irvine}},\ }\bibfield  {title} {\bibinfo {title}
  {{Topological mechanics of gyroscopic metamaterials}},\ }\bibfield  {journal}
  {\bibinfo  {journal} {Proc. Natl. Acad. Sci. U.S.A.}\ }\textbf {\bibinfo
  {volume} {112}},\ \href {https://doi.org/10.1073/pnas.1507413112}
  {10.1073/pnas.1507413112} (\bibinfo {year} {2015}),\ \Eprint
  {https://arxiv.org/abs/1504.03362} {arXiv:1504.03362} \BibitemShut {NoStop}%
\bibitem [{\citenamefont {Wang}\ \emph
  {et~al.}(2015{\natexlab{a}})\citenamefont {Wang}, \citenamefont {Lu},\ and\
  \citenamefont {Bertoldi}}]{Wang2015}%
  \BibitemOpen
  \bibfield  {author} {\bibinfo {author} {\bibfnamefont {P.}~\bibnamefont
  {Wang}}, \bibinfo {author} {\bibfnamefont {L.}~\bibnamefont {Lu}}, and\
  \bibinfo {author} {\bibfnamefont {K.}~\bibnamefont {Bertoldi}},\ }\bibfield
  {title} {\bibinfo {title} {{Topological Phononic Crystals with One-Way
  Elastic Edge Waves}},\ }\href
  {https://doi.org/10.1103/PhysRevLett.115.104302} {\bibfield  {journal}
  {\bibinfo  {journal} {Phys. Rev. Lett.}\ }\textbf {\bibinfo {volume} {115}},\
  \bibinfo {pages} {104302} (\bibinfo {year} {2015}{\natexlab{a}})},\ \Eprint
  {https://arxiv.org/abs/1504.01374v1} {arXiv:1504.01374v1} \BibitemShut
  {NoStop}%
\bibitem [{\citenamefont {Men}\ \emph {et~al.}(2014)\citenamefont {Men},
  \citenamefont {Lee}, \citenamefont {Freund}, \citenamefont {Peraire},\ and\
  \citenamefont {Johnson}}]{Men2014}%
  \BibitemOpen
  \bibfield  {author} {\bibinfo {author} {\bibfnamefont {H.}~\bibnamefont
  {Men}}, \bibinfo {author} {\bibfnamefont {K.~Y.~K.}\ \bibnamefont {Lee}},
  \bibinfo {author} {\bibfnamefont {R.~M.}\ \bibnamefont {Freund}}, \bibinfo
  {author} {\bibfnamefont {J.}~\bibnamefont {Peraire}}, and\ \bibinfo {author}
  {\bibfnamefont {S.~G.}\ \bibnamefont {Johnson}},\ }\bibfield  {title}
  {\bibinfo {title} {{Robust topology optimization of three-dimensional
  photonic-crystal band-gap structures}},\ }\href
  {https://doi.org/10.1364/OE.22.022632} {\bibfield  {journal} {\bibinfo
  {journal} {Opt. Express}\ }\textbf {\bibinfo {volume} {22}},\ \bibinfo
  {pages} {263} (\bibinfo {year} {2014})},\ \Eprint
  {https://arxiv.org/abs/1405.4350} {arXiv:1405.4350} \BibitemShut {NoStop}%
\bibitem [{\citenamefont {{Olsson III}}\ and\ \citenamefont
  {El-Kady}(2009)}]{OlssonIII2009}%
  \BibitemOpen
  \bibfield  {author} {\bibinfo {author} {\bibfnamefont {R.~H.}\ \bibnamefont
  {{Olsson III}}}and\ \bibinfo {author} {\bibfnamefont {I.}~\bibnamefont
  {El-Kady}},\ }\bibfield  {title} {\bibinfo {title} {{Microfabricated phononic
  crystal devices and applications}},\ }\href
  {https://doi.org/10.1088/0957-0233/20/1/012002} {\bibfield  {journal}
  {\bibinfo  {journal} {Meas. Sci. Technol}\ }\textbf {\bibinfo {volume}
  {20}},\ \bibinfo {pages} {012002} (\bibinfo {year} {2009})}\BibitemShut
  {NoStop}%
\bibitem [{\citenamefont {Goffaux}\ and\ \citenamefont
  {Vigneron}(2001)}]{Goffaux2001}%
  \BibitemOpen
  \bibfield  {author} {\bibinfo {author} {\bibfnamefont {C.}~\bibnamefont
  {Goffaux}}and\ \bibinfo {author} {\bibfnamefont {J.~P.}\ \bibnamefont
  {Vigneron}},\ }\bibfield  {title} {\bibinfo {title} {{Theoretical study of a
  tunable phononic band gap system}},\ }\href
  {https://doi.org/10.1103/PhysRevB.64.075118} {\bibfield  {journal} {\bibinfo
  {journal} {Phys. Rev. B}\ }\textbf {\bibinfo {volume} {64}},\ \bibinfo
  {pages} {075118} (\bibinfo {year} {2001})}\BibitemShut {NoStop}%
\bibitem [{\citenamefont {Agarwala}\ and\ \citenamefont
  {Shenoy}(2017)}]{Agarwala2017}%
  \BibitemOpen
  \bibfield  {author} {\bibinfo {author} {\bibfnamefont {A.}~\bibnamefont
  {Agarwala}}and\ \bibinfo {author} {\bibfnamefont {V.~B.}\ \bibnamefont
  {Shenoy}},\ }\bibfield  {title} {\bibinfo {title} {{Topological Insulators in
  Amorphous Systems}},\ }\href {https://doi.org/10.1103/PhysRevLett.118.236402}
  {\bibfield  {journal} {\bibinfo  {journal} {Phys. Rev. Lett.}\ }\textbf
  {\bibinfo {volume} {118}},\ \bibinfo {pages} {236402} (\bibinfo {year}
  {2017})},\ \Eprint {https://arxiv.org/abs/1701.00374} {arXiv:1701.00374}
  \BibitemShut {NoStop}%
\bibitem [{\citenamefont {Mitchell}\ \emph
  {et~al.}(2018{\natexlab{a}})\citenamefont {Mitchell}, \citenamefont {Nash},\
  and\ \citenamefont {Irvine}}]{Mitchell2018}%
  \BibitemOpen
  \bibfield  {author} {\bibinfo {author} {\bibfnamefont {N.~P.}\ \bibnamefont
  {Mitchell}}, \bibinfo {author} {\bibfnamefont {L.~M.}\ \bibnamefont {Nash}},
  and\ \bibinfo {author} {\bibfnamefont {W.~T.~M.}\ \bibnamefont {Irvine}},\
  }\bibfield  {title} {\bibinfo {title} {{Realization of a topological phase
  transition in a gyroscopic lattice}},\ }\href
  {https://doi.org/10.1103/PhysRevB.97.100302} {\bibfield  {journal} {\bibinfo
  {journal} {Phys. Rev. B}\ }\textbf {\bibinfo {volume} {97}},\ \bibinfo
  {pages} {100302} (\bibinfo {year} {2018}{\natexlab{a}})},\ \Eprint
  {https://arxiv.org/abs/1711.02433} {arXiv:1711.02433} \BibitemShut {NoStop}%
\bibitem [{\citenamefont {Yan}\ \emph {et~al.}(2017)\citenamefont {Yan},
  \citenamefont {Ravasio}, \citenamefont {Brito},\ and\ \citenamefont
  {Wyart}}]{Yan2017}%
  \BibitemOpen
  \bibfield  {author} {\bibinfo {author} {\bibfnamefont {L.}~\bibnamefont
  {Yan}}, \bibinfo {author} {\bibfnamefont {R.}~\bibnamefont {Ravasio}},
  \bibinfo {author} {\bibfnamefont {C.}~\bibnamefont {Brito}}, and\ \bibinfo
  {author} {\bibfnamefont {M.}~\bibnamefont {Wyart}},\ }\bibfield  {title}
  {\bibinfo {title} {{Architecture and coevolution of allosteric materials}},\
  }\href {https://doi.org/10.1073/pnas.1615536114} {\bibfield  {journal}
  {\bibinfo  {journal} {Proc. Natl. Acad. Sci. U.S.A.}\ }\textbf {\bibinfo
  {volume} {114}},\ \bibinfo {pages} {2526} (\bibinfo {year} {2017})},\ \Eprint
  {https://arxiv.org/abs/1609.03951} {arXiv:1609.03951} \BibitemShut {NoStop}%
\bibitem [{\citenamefont {Reid}\ \emph {et~al.}(2018)\citenamefont {Reid},
  \citenamefont {Pashine}, \citenamefont {Wozniak}, \citenamefont {Jaeger},
  \citenamefont {Liu}, \citenamefont {Nagel},\ and\ \citenamefont
  {de~Pablo}}]{Reid2017}%
  \BibitemOpen
  \bibfield  {author} {\bibinfo {author} {\bibfnamefont {D.~R.}\ \bibnamefont
  {Reid}}, \bibinfo {author} {\bibfnamefont {N.}~\bibnamefont {Pashine}},
  \bibinfo {author} {\bibfnamefont {J.~M.}\ \bibnamefont {Wozniak}}, \bibinfo
  {author} {\bibfnamefont {H.~M.}\ \bibnamefont {Jaeger}}, \bibinfo {author}
  {\bibfnamefont {A.~J.}\ \bibnamefont {Liu}}, \bibinfo {author} {\bibfnamefont
  {S.~R.}\ \bibnamefont {Nagel}}, and\ \bibinfo {author} {\bibfnamefont
  {J.~J.}\ \bibnamefont {de~Pablo}},\ }\bibfield  {title} {\bibinfo {title}
  {{Auxetic metamaterials from disordered networks}},\ }\href
  {https://doi.org/10.1073/pnas.1717442115} {\bibfield  {journal} {\bibinfo
  {journal} {Proc. Natl. Acad. Sci. U.S.A.}\ }\textbf {\bibinfo {volume}
  {115}},\ \bibinfo {pages} {E1384} (\bibinfo {year} {2018})},\ \Eprint
  {https://arxiv.org/abs/1710.02493} {arXiv:1710.02493} \BibitemShut {NoStop}%
\bibitem [{\citenamefont {Flechsig}(2017)}]{Flechsig2017}%
  \BibitemOpen
  \bibfield  {author} {\bibinfo {author} {\bibfnamefont {H.}~\bibnamefont
  {Flechsig}},\ }\bibfield  {title} {\bibinfo {title} {{Design of Elastic
  Networks with Evolutionary Optimized Long-Range Communication as Mechanical
  Models of Allosteric Proteins}},\ }\href
  {https://doi.org/10.1016/j.bpj.2017.06.043} {\bibfield  {journal} {\bibinfo
  {journal} {Biophys. J.}\ }\textbf {\bibinfo {volume} {113}},\ \bibinfo
  {pages} {558} (\bibinfo {year} {2017})},\ \Eprint
  {https://arxiv.org/abs/1702.08317} {arXiv:1702.08317} \BibitemShut {NoStop}%
\bibitem [{\citenamefont {Rocks}\ \emph {et~al.}(2017)\citenamefont {Rocks},
  \citenamefont {Pashine}, \citenamefont {Bischofberger}, \citenamefont
  {Goodrich}, \citenamefont {Liu},\ and\ \citenamefont {Nagel}}]{Rocks2017}%
  \BibitemOpen
  \bibfield  {author} {\bibinfo {author} {\bibfnamefont {J.~W.}\ \bibnamefont
  {Rocks}}, \bibinfo {author} {\bibfnamefont {N.}~\bibnamefont {Pashine}},
  \bibinfo {author} {\bibfnamefont {I.}~\bibnamefont {Bischofberger}}, \bibinfo
  {author} {\bibfnamefont {C.~P.}\ \bibnamefont {Goodrich}}, \bibinfo {author}
  {\bibfnamefont {A.~J.}\ \bibnamefont {Liu}}, and\ \bibinfo {author}
  {\bibfnamefont {S.~R.}\ \bibnamefont {Nagel}},\ }\bibfield  {title} {\bibinfo
  {title} {{Designing allostery-inspired response in mechanical networks}},\
  }\href {https://doi.org/10.1073/pnas.1612139114} {\bibfield  {journal}
  {\bibinfo  {journal} {Proc. Natl. Acad. Sci. U.S.A.}\ }\textbf {\bibinfo
  {volume} {114}},\ \bibinfo {pages} {2520} (\bibinfo {year} {2017})},\ \Eprint
  {https://arxiv.org/abs/1607.08562v1} {arXiv:1607.08562v1} \BibitemShut
  {NoStop}%
\bibitem [{\citenamefont {Goodrich}\ \emph {et~al.}(2015)\citenamefont
  {Goodrich}, \citenamefont {Liu},\ and\ \citenamefont {Nagel}}]{Goodrich2015}%
  \BibitemOpen
  \bibfield  {author} {\bibinfo {author} {\bibfnamefont {C.~P.}\ \bibnamefont
  {Goodrich}}, \bibinfo {author} {\bibfnamefont {A.~J.}\ \bibnamefont {Liu}},
  and\ \bibinfo {author} {\bibfnamefont {S.~R.}\ \bibnamefont {Nagel}},\
  }\bibfield  {title} {\bibinfo {title} {{The Principle of Independent
  Bond-Level Response: Tuning by Pruning to Exploit Disorder for Global
  Behavior}},\ }\href {https://doi.org/10.1103/PhysRevLett.114.225501}
  {\bibfield  {journal} {\bibinfo  {journal} {Phys. Rev. Lett.}\ }\textbf
  {\bibinfo {volume} {114}},\ \bibinfo {pages} {8} (\bibinfo {year} {2015})},\
  \Eprint {https://arxiv.org/abs/1502.02953} {arXiv:1502.02953} \BibitemShut
  {NoStop}%
\bibitem [{\citenamefont {Rocks}\ \emph {et~al.}(2019)\citenamefont {Rocks},
  \citenamefont {Ronellenfitsch}, \citenamefont {Liu}, \citenamefont {Nagel},\
  and\ \citenamefont {Katifori}}]{Rocks2019}%
  \BibitemOpen
  \bibfield  {author} {\bibinfo {author} {\bibfnamefont {J.~W.}\ \bibnamefont
  {Rocks}}, \bibinfo {author} {\bibfnamefont {H.}~\bibnamefont
  {Ronellenfitsch}}, \bibinfo {author} {\bibfnamefont {A.~J.}\ \bibnamefont
  {Liu}}, \bibinfo {author} {\bibfnamefont {S.~R.}\ \bibnamefont {Nagel}}, and\
  \bibinfo {author} {\bibfnamefont {E.}~\bibnamefont {Katifori}},\ }\bibfield
  {title} {\bibinfo {title} {{Limits of multifunctionality in tunable
  networks}},\ }\href {https://doi.org/10.1073/pnas.1806790116} {\bibfield
  {journal} {\bibinfo  {journal} {Proc. Natl. Acad. Sci. U.S.A.}\ }\textbf
  {\bibinfo {volume} {11}},\ \bibinfo {pages} {201806790} (\bibinfo {year}
  {2019})},\ \Eprint {https://arxiv.org/abs/1805.00504} {arXiv:1805.00504}
  \BibitemShut {NoStop}%
\bibitem [{\citenamefont {Jensen}(2003)}]{Jensen2003}%
  \BibitemOpen
  \bibfield  {author} {\bibinfo {author} {\bibfnamefont {J.~S.}\ \bibnamefont
  {Jensen}},\ }\bibfield  {title} {\bibinfo {title} {{Phononic band gaps and
  vibrations in one- and two-dimensional mass--spring structures}},\ }\href
  {https://doi.org/10.1016/S0022-460X(02)01629-2} {\bibfield  {journal}
  {\bibinfo  {journal} {J. Sound. Vib.}\ }\textbf {\bibinfo {volume} {266}},\
  \bibinfo {pages} {1053} (\bibinfo {year} {2003})}\BibitemShut {NoStop}%
\bibitem [{\citenamefont {Lubensky}\ \emph {et~al.}(2015)\citenamefont
  {Lubensky}, \citenamefont {Kane}, \citenamefont {Mao}, \citenamefont
  {Souslov},\ and\ \citenamefont {Sun}}]{Lubensky2015}%
  \BibitemOpen
  \bibfield  {author} {\bibinfo {author} {\bibfnamefont {T.~C.}\ \bibnamefont
  {Lubensky}}, \bibinfo {author} {\bibfnamefont {C.~L.}\ \bibnamefont {Kane}},
  \bibinfo {author} {\bibfnamefont {X.}~\bibnamefont {Mao}}, \bibinfo {author}
  {\bibfnamefont {A.}~\bibnamefont {Souslov}}, and\ \bibinfo {author}
  {\bibfnamefont {K.}~\bibnamefont {Sun}},\ }\bibfield  {title} {\bibinfo
  {title} {{Phonons and elasticity in critically coordinated lattices}},\
  }\href {https://doi.org/10.1088/0034-4885/78/7/073901} {\bibfield  {journal}
  {\bibinfo  {journal} {Rep. Prog. Phys.}\ }\textbf {\bibinfo {volume} {78}},\
  \bibinfo {pages} {073901} (\bibinfo {year} {2015})},\ \Eprint
  {https://arxiv.org/abs/1503.01324v1} {arXiv:1503.01324v1} \BibitemShut
  {NoStop}%
\bibitem [{\citenamefont {Men}\ \emph {et~al.}(2010)\citenamefont {Men},
  \citenamefont {Nguyen}, \citenamefont {Freund}, \citenamefont {Parrilo},\
  and\ \citenamefont {Peraire}}]{Men2010}%
  \BibitemOpen
  \bibfield  {author} {\bibinfo {author} {\bibfnamefont {H.}~\bibnamefont
  {Men}}, \bibinfo {author} {\bibfnamefont {N.}~\bibnamefont {Nguyen}},
  \bibinfo {author} {\bibfnamefont {R.}~\bibnamefont {Freund}}, \bibinfo
  {author} {\bibfnamefont {P.}~\bibnamefont {Parrilo}}, and\ \bibinfo {author}
  {\bibfnamefont {J.}~\bibnamefont {Peraire}},\ }\bibfield  {title} {\bibinfo
  {title} {{Bandgap optimization of two-dimensional photonic crystals using
  semidefinite programming and subspace methods}},\ }\href
  {https://doi.org/10.1016/j.jcp.2010.01.023} {\bibfield  {journal} {\bibinfo
  {journal} {J. Comput. Phys.}\ }\textbf {\bibinfo {volume} {229}},\ \bibinfo
  {pages} {3706} (\bibinfo {year} {2010})}\BibitemShut {NoStop}%
\bibitem [{\citenamefont {Ronellenfitsch}\ \emph {et~al.}(2018)\citenamefont
  {Ronellenfitsch}, \citenamefont {Dunkel},\ and\ \citenamefont
  {Wilczek}}]{Ronellenfitsch2018a}%
  \BibitemOpen
  \bibfield  {author} {\bibinfo {author} {\bibfnamefont {H.}~\bibnamefont
  {Ronellenfitsch}}, \bibinfo {author} {\bibfnamefont {J.}~\bibnamefont
  {Dunkel}}, and\ \bibinfo {author} {\bibfnamefont {M.}~\bibnamefont
  {Wilczek}},\ }\bibfield  {title} {\bibinfo {title} {{Optimal Noise-Canceling
  Networks}},\ }\href {https://doi.org/10.1103/PhysRevLett.121.208301}
  {\bibfield  {journal} {\bibinfo  {journal} {Phys. Rev. Lett.}\ }\textbf
  {\bibinfo {volume} {121}},\ \bibinfo {pages} {208301} (\bibinfo {year}
  {2018})},\ \Eprint {https://arxiv.org/abs/1807.08376v2} {arXiv:1807.08376v2}
  \BibitemShut {NoStop}%
\bibitem [{\citenamefont {Bertoldi}\ and\ \citenamefont
  {Boyce}(2008)}]{Bertoldi2008}%
  \BibitemOpen
  \bibfield  {author} {\bibinfo {author} {\bibfnamefont {K.}~\bibnamefont
  {Bertoldi}}and\ \bibinfo {author} {\bibfnamefont {M.~C.}\ \bibnamefont
  {Boyce}},\ }\bibfield  {title} {\bibinfo {title} {{Mechanically triggered
  transformations of phononic band gaps in periodic elastomeric structures}},\
  }\href {https://doi.org/10.1103/PhysRevB.77.052105} {\bibfield  {journal}
  {\bibinfo  {journal} {Phys. Rev. B}\ }\textbf {\bibinfo {volume} {77}},\
  \bibinfo {pages} {052105} (\bibinfo {year} {2008})}\BibitemShut {NoStop}%
\bibitem [{\citenamefont {Gui}\ \emph {et~al.}(2008)\citenamefont {Gui},
  \citenamefont {Li},\ and\ \citenamefont {Zhong}}]{Gui2008}%
  \BibitemOpen
  \bibfield  {author} {\bibinfo {author} {\bibfnamefont {G.}~\bibnamefont
  {Gui}}, \bibinfo {author} {\bibfnamefont {J.}~\bibnamefont {Li}}, and\
  \bibinfo {author} {\bibfnamefont {J.}~\bibnamefont {Zhong}},\ }\bibfield
  {title} {\bibinfo {title} {{Band structure engineering of graphene by strain:
  First-principles calculations}},\ }\href
  {https://doi.org/10.1103/PhysRevB.78.075435} {\bibfield  {journal} {\bibinfo
  {journal} {Phys. Rev. B}\ }\textbf {\bibinfo {volume} {78}},\ \bibinfo
  {pages} {075435} (\bibinfo {year} {2008})},\ \Eprint
  {https://arxiv.org/abs/0903.1702} {arXiv:0903.1702} \BibitemShut {NoStop}%
\bibitem [{\citenamefont {Wang}\ and\ \citenamefont
  {Bertoldi}(2012)}]{Wang2012}%
  \BibitemOpen
  \bibfield  {author} {\bibinfo {author} {\bibfnamefont {L.}~\bibnamefont
  {Wang}}and\ \bibinfo {author} {\bibfnamefont {K.}~\bibnamefont {Bertoldi}},\
  }\bibfield  {title} {\bibinfo {title} {{Mechanically tunable phononic band
  gaps in three-dimensional periodic elastomeric structures}},\ }\href
  {https://doi.org/10.1016/j.ijsolstr.2012.05.008} {\bibfield  {journal}
  {\bibinfo  {journal} {Int. J. Solids. Struct.}\ }\textbf {\bibinfo {volume}
  {49}},\ \bibinfo {pages} {2881} (\bibinfo {year} {2012})}\BibitemShut
  {NoStop}%
\bibitem [{\citenamefont {Miettinen}(1998)}]{Miettinen1998}%
  \BibitemOpen
  \bibfield  {author} {\bibinfo {author} {\bibfnamefont {K.}~\bibnamefont
  {Miettinen}},\ }\href {https://doi.org/10.1007/978-1-4615-5563-6} {\emph
  {\bibinfo {title} {{Nonlinear Multiobjective Optimization}}}},\ \bibinfo
  {series} {International Series in Operations Research {\&} Management
  Science}, Vol.~\bibinfo {volume} {12}\ (\bibinfo  {publisher} {Springer US},\
  \bibinfo {address} {Boston, MA},\ \bibinfo {year} {1998})\ p.\ \bibinfo
  {pages} {404}\BibitemShut {NoStop}%
\bibitem [{\citenamefont {Audoly}\ and\ \citenamefont
  {Pomeau}(2010)}]{Audoly2010}%
  \BibitemOpen
  \bibfield  {author} {\bibinfo {author} {\bibfnamefont {B.}~\bibnamefont
  {Audoly}}and\ \bibinfo {author} {\bibfnamefont {Y.}~\bibnamefont {Pomeau}},\
  }\href@noop {} {\emph {\bibinfo {title} {{Elasticity and Geometry}}}}\
  (\bibinfo  {publisher} {Oxford University Press},\ \bibinfo {address}
  {Oxford},\ \bibinfo {year} {2010})\ p.\ \bibinfo {pages} {586}\BibitemShut
  {NoStop}%
\bibitem [{\citenamefont {Bergou}\ \emph {et~al.}(2008)\citenamefont {Bergou},
  \citenamefont {Wardetzky}, \citenamefont {Robinson}, \citenamefont {Audoly},\
  and\ \citenamefont {Grinspun}}]{Bergou2008}%
  \BibitemOpen
  \bibfield  {author} {\bibinfo {author} {\bibfnamefont {M.}~\bibnamefont
  {Bergou}}, \bibinfo {author} {\bibfnamefont {M.}~\bibnamefont {Wardetzky}},
  \bibinfo {author} {\bibfnamefont {S.}~\bibnamefont {Robinson}}, \bibinfo
  {author} {\bibfnamefont {B.}~\bibnamefont {Audoly}}, and\ \bibinfo {author}
  {\bibfnamefont {E.}~\bibnamefont {Grinspun}},\ }\bibfield  {title} {\bibinfo
  {title} {Discrete elastic rods},\ }\href
  {https://doi.org/10.1145/1360612.1360662} {\bibfield  {journal} {\bibinfo
  {journal} {ACM Trans. Graph.}\ }\textbf {\bibinfo {volume} {27}},\ \bibinfo
  {pages} {63:1} (\bibinfo {year} {2008})}\BibitemShut {NoStop}%
\bibitem [{\citenamefont {Kane}\ and\ \citenamefont
  {Lubensky}(2014)}]{Kane2014}%
  \BibitemOpen
  \bibfield  {author} {\bibinfo {author} {\bibfnamefont {C.~L.}\ \bibnamefont
  {Kane}}and\ \bibinfo {author} {\bibfnamefont {T.~C.}\ \bibnamefont
  {Lubensky}},\ }\bibfield  {title} {\bibinfo {title} {{Topological boundary
  modes in isostatic lattices}},\ }\href {https://doi.org/10.1038/nphys2835}
  {\bibfield  {journal} {\bibinfo  {journal} {Nat. Phys.}\ }\textbf {\bibinfo
  {volume} {10}},\ \bibinfo {pages} {39} (\bibinfo {year} {2014})},\ \Eprint
  {https://arxiv.org/abs/1308.0554} {arXiv:1308.0554} \BibitemShut {NoStop}%
\bibitem [{\citenamefont {Souslov}\ \emph {et~al.}(2017)\citenamefont
  {Souslov}, \citenamefont {van Zuiden}, \citenamefont {Bartolo},\ and\
  \citenamefont {Vitelli}}]{Souslov2016}%
  \BibitemOpen
  \bibfield  {author} {\bibinfo {author} {\bibfnamefont {A.}~\bibnamefont
  {Souslov}}, \bibinfo {author} {\bibfnamefont {B.~C.}\ \bibnamefont {van
  Zuiden}}, \bibinfo {author} {\bibfnamefont {D.}~\bibnamefont {Bartolo}}, and\
  \bibinfo {author} {\bibfnamefont {V.}~\bibnamefont {Vitelli}},\ }\bibfield
  {title} {\bibinfo {title} {{Topological sound in active-liquid
  metamaterials}},\ }\href {https://doi.org/10.1038/nphys4193} {\bibfield
  {journal} {\bibinfo  {journal} {Nat. Phys.}\ }\textbf {\bibinfo {volume}
  {13}},\ \bibinfo {pages} {1091} (\bibinfo {year} {2017})}\BibitemShut
  {NoStop}%
\bibitem [{\citenamefont {Mitchell}\ \emph
  {et~al.}(2018{\natexlab{b}})\citenamefont {Mitchell}, \citenamefont {Nash},
  \citenamefont {Hexner}, \citenamefont {Turner},\ and\ \citenamefont
  {Irvine}}]{Mitchell2018b}%
  \BibitemOpen
  \bibfield  {author} {\bibinfo {author} {\bibfnamefont {N.~P.}\ \bibnamefont
  {Mitchell}}, \bibinfo {author} {\bibfnamefont {L.~M.}\ \bibnamefont {Nash}},
  \bibinfo {author} {\bibfnamefont {D.}~\bibnamefont {Hexner}}, \bibinfo
  {author} {\bibfnamefont {A.~M.}\ \bibnamefont {Turner}}, and\ \bibinfo
  {author} {\bibfnamefont {W.~T.~M.}\ \bibnamefont {Irvine}},\ }\bibfield
  {title} {\bibinfo {title} {{Amorphous topological insulators constructed from
  random point sets}},\ }\href {https://doi.org/10.1038/s41567-017-0024-5}
  {\bibfield  {journal} {\bibinfo  {journal} {Nat. Phys.}\ }\textbf {\bibinfo
  {volume} {14}},\ \bibinfo {pages} {380} (\bibinfo {year}
  {2018}{\natexlab{b}})}\BibitemShut {NoStop}%
\bibitem [{\citenamefont {Wang}\ \emph
  {et~al.}(2015{\natexlab{b}})\citenamefont {Wang}, \citenamefont {Luan},\ and\
  \citenamefont {Zhang}}]{Wang2015c}%
  \BibitemOpen
  \bibfield  {author} {\bibinfo {author} {\bibfnamefont {Y.-T.}\ \bibnamefont
  {Wang}}, \bibinfo {author} {\bibfnamefont {P.-G.}\ \bibnamefont {Luan}}, and\
  \bibinfo {author} {\bibfnamefont {S.}~\bibnamefont {Zhang}},\ }\bibfield
  {title} {\bibinfo {title} {{Coriolis force induced topological order for
  classical mechanical vibrations}},\ }\href
  {https://doi.org/10.1088/1367-2630/17/7/073031} {\bibfield  {journal}
  {\bibinfo  {journal} {New J. Phys.}\ }\textbf {\bibinfo {volume} {17}},\
  \bibinfo {pages} {073031} (\bibinfo {year} {2015}{\natexlab{b}})},\ \Eprint
  {https://arxiv.org/abs/1411.2806} {arXiv:1411.2806} \BibitemShut {NoStop}%
\bibitem [{\citenamefont {Ricouvier}\ \emph {et~al.}(2019)\citenamefont
  {Ricouvier}, \citenamefont {Tabeling},\ and\ \citenamefont
  {Yazhgur}}]{Ricouvier2019}%
  \BibitemOpen
  \bibfield  {author} {\bibinfo {author} {\bibfnamefont {J.}~\bibnamefont
  {Ricouvier}}, \bibinfo {author} {\bibfnamefont {P.}~\bibnamefont {Tabeling}},
  and\ \bibinfo {author} {\bibfnamefont {P.}~\bibnamefont {Yazhgur}},\
  }\bibfield  {title} {\bibinfo {title} {{Foam as a self-assembling amorphous
  photonic band gap material}},\ }\href
  {https://doi.org/10.1073/pnas.1820526116} {\bibfield  {journal} {\bibinfo
  {journal} {Proc. Natl. Acad. Sci. U.S.A.}\ ,\ \bibinfo {pages} {201820526}}
  (\bibinfo {year} {2019})}\BibitemShut {NoStop}%
\bibitem [{\citenamefont {Bathe}\ and\ \citenamefont
  {Rothemund}(2017)}]{Bathe2017}%
  \BibitemOpen
  \bibfield  {author} {\bibinfo {author} {\bibfnamefont {M.}~\bibnamefont
  {Bathe}}and\ \bibinfo {author} {\bibfnamefont {P.~W.}\ \bibnamefont
  {Rothemund}},\ }\bibfield  {title} {\bibinfo {title} {{DNA} nanotechnology: A
  foundation for programmable nanoscale materials},\ }\href
  {https://doi.org/10.1557/mrs.2017.279} {\bibfield  {journal} {\bibinfo
  {journal} {MRS Bulletin}\ }\textbf {\bibinfo {volume} {42}},\ \bibinfo
  {pages} {882–888} (\bibinfo {year} {2017})}\BibitemShut {NoStop}%
\bibitem [{\citenamefont {Sun}\ \emph {et~al.}(2014)\citenamefont {Sun},
  \citenamefont {Boulais}, \citenamefont {Hakobyan}, \citenamefont {Wang},
  \citenamefont {Guan}, \citenamefont {Bathe},\ and\ \citenamefont
  {Yin}}]{Sun2014}%
  \BibitemOpen
  \bibfield  {author} {\bibinfo {author} {\bibfnamefont {W.}~\bibnamefont
  {Sun}}, \bibinfo {author} {\bibfnamefont {E.}~\bibnamefont {Boulais}},
  \bibinfo {author} {\bibfnamefont {Y.}~\bibnamefont {Hakobyan}}, \bibinfo
  {author} {\bibfnamefont {W.~L.}\ \bibnamefont {Wang}}, \bibinfo {author}
  {\bibfnamefont {A.}~\bibnamefont {Guan}}, \bibinfo {author} {\bibfnamefont
  {M.}~\bibnamefont {Bathe}}, and\ \bibinfo {author} {\bibfnamefont
  {P.}~\bibnamefont {Yin}},\ }\bibfield  {title} {\bibinfo {title} {{Casting
  inorganic structures with DNA molds}},\ }\href
  {https://doi.org/10.1126/science.1258361} {\bibfield  {journal} {\bibinfo
  {journal} {Science}\ }\textbf {\bibinfo {volume} {346}},\ \bibinfo {pages}
  {1258361} (\bibinfo {year} {2014})}\BibitemShut {NoStop}%
\bibitem [{\citenamefont {Lee}\ \emph {et~al.}(2018)\citenamefont {Lee},
  \citenamefont {Imhof}, \citenamefont {Berger}, \citenamefont {Bayer},
  \citenamefont {Brehm}, \citenamefont {Molenkamp}, \citenamefont {Kiessling},\
  and\ \citenamefont {Thomale}}]{Lee2017}%
  \BibitemOpen
  \bibfield  {author} {\bibinfo {author} {\bibfnamefont {C.~H.}\ \bibnamefont
  {Lee}}, \bibinfo {author} {\bibfnamefont {S.}~\bibnamefont {Imhof}}, \bibinfo
  {author} {\bibfnamefont {C.}~\bibnamefont {Berger}}, \bibinfo {author}
  {\bibfnamefont {F.}~\bibnamefont {Bayer}}, \bibinfo {author} {\bibfnamefont
  {J.}~\bibnamefont {Brehm}}, \bibinfo {author} {\bibfnamefont {L.~W.}\
  \bibnamefont {Molenkamp}}, \bibinfo {author} {\bibfnamefont {T.}~\bibnamefont
  {Kiessling}}, and\ \bibinfo {author} {\bibfnamefont {R.}~\bibnamefont
  {Thomale}},\ }\bibfield  {title} {\bibinfo {title} {{Topolectrical
  Circuits}},\ }\href {https://doi.org/10.1038/s42005-018-0035-2} {\bibfield
  {journal} {\bibinfo  {journal} {Commun. Phys.}\ }\textbf {\bibinfo {volume}
  {1}},\ \bibinfo {pages} {39} (\bibinfo {year} {2018})},\ \Eprint
  {https://arxiv.org/abs/1705.01077} {arXiv:1705.01077} \BibitemShut {NoStop}%
\bibitem [{\citenamefont {Kotwal}\ \emph {et~al.}(2019)\citenamefont {Kotwal},
  \citenamefont {Ronellenfitsch}, \citenamefont {Moseley},\ and\ \citenamefont
  {Dunkel}}]{Kotwal2019}%
  \BibitemOpen
  \bibfield  {author} {\bibinfo {author} {\bibfnamefont {T.}~\bibnamefont
  {Kotwal}}, \bibinfo {author} {\bibfnamefont {H.}~\bibnamefont
  {Ronellenfitsch}}, \bibinfo {author} {\bibfnamefont {F.}~\bibnamefont
  {Moseley}}, and\ \bibinfo {author} {\bibfnamefont {J.}~\bibnamefont
  {Dunkel}},\ }\bibfield  {title} {\bibinfo {title} {{Active topolectrical
  circuits}},\ }\href@noop {} {\bibfield  {journal} {\bibinfo  {journal} {arXiv
  preprint}\ } (\bibinfo {year} {2019})},\ \Eprint
  {https://arxiv.org/abs/1903.10130v1} {arXiv:1903.10130v1} \BibitemShut
  {NoStop}%
\bibitem [{\citenamefont {Byrd}\ \emph {et~al.}(1995)\citenamefont {Byrd},
  \citenamefont {Lu}, \citenamefont {Nocedal},\ and\ \citenamefont
  {Zhu}}]{Byrd1995}%
  \BibitemOpen
  \bibfield  {author} {\bibinfo {author} {\bibfnamefont {R.~H.}\ \bibnamefont
  {Byrd}}, \bibinfo {author} {\bibfnamefont {P.}~\bibnamefont {Lu}}, \bibinfo
  {author} {\bibfnamefont {J.}~\bibnamefont {Nocedal}}, and\ \bibinfo {author}
  {\bibfnamefont {C.}~\bibnamefont {Zhu}},\ }\bibfield  {title} {\bibinfo
  {title} {{A Limited Memory Algorithm for Bound Constrained Optimization}},\
  }\href {https://doi.org/10.1137/0916069} {\bibfield  {journal} {\bibinfo
  {journal} {SIAM J. Sci. Comput.}\ }\textbf {\bibinfo {volume} {16}},\
  \bibinfo {pages} {1190} (\bibinfo {year} {1995})}\BibitemShut {NoStop}%
\end{thebibliography}%

\newpage
\onecolumngrid
\section{Supplementary Material}

\beginsupplement

\section{Linear approximation for balls-and-springs networks}
We consider networks of point masses connected by springs.
The elastic energy of such a single spring between two masses is given by
\begin{align}
  V = \frac{1}{2} k_{12} \left( |\vec{x}_1 - \vec{x}_2| - \ell \right)^2,
\end{align}
where $\vec x_i$ is the position of mass $i$, $k_{12}$  the spring constant/stiffness,
and $\ell$  the equilibrium length.
We introduce the (small) displacements $\vec u_i$ by $\vec x_i = \vec
x_i^{(0)} + \vec u_i$,
where $\vec x_i^{(0)}$ is the rest position of mass $i$.
Then, neglecting terms of $\mathcal{O}(\vec u^2)$, we find for the elastic energy
\begin{align}
  V \approx \frac{1}{2} k_{12} \left( \hat b_{12}\cdot (\vec u_1 - \vec u_2) \right)^2,
\end{align}
where $\hat b_{12} = (\vec x_1^{(0)} - \vec x_2^{(0)})/\ell$ is the normal vector along
the spring at equilibrium.
The quantity $\hat b_{12}\cdot (\vec u_1 - \vec u_2)$ is referred to as the elongation from now on.

Generalizing to a network of springs, we can write the elastic energy as
\begin{align}
  V = \frac{1}{2} \mathbf{e}^\top k \mathbf{e}
  = \frac{1}{2} \mathbf{u}^\top K \mathbf{u}, \qquad K = QkQ^\top,
\end{align}
where $\mathbf e$ is the vector of bond elongations, $k$  a diagonal matrix of
spring constants, $\mathbf u$ the vector of displacement components,
$K$ the stiffness matrix, and $Q$ the equilibrium matrix which maps
$\mathbf f = Q \mathbf t$, where $\mathbf f$ is the vector of net force
(or external load) components
at each node and $\mathbf t$ is the vector of spring tensions.
There is also the compatibility matrix $C = Q^\top$ which maps elongations to
displacements, $C \mathbf{u} = \mathbf e$.
The compatibility/equilibrium matrix encodes the relative spatial
relationships between the nodes at rest.

If the oriented incidence matrix $B$ of the network is known,
then in $d$ dimensions,
\begin{align}
  Q = \begin{pmatrix}
  B * \hat b_{x_1} \\
  B * \hat b_{x_2} \\
  \vdots \\
  B * \hat b_{x_d}
\end{pmatrix}.
\end{align}
Here, we define $(B * v)_{ij} = B_{ij} v_j$ (no sum implied), and
we have stacked $d$ copies of the incidence matrix and weighted each
edge by the appropriate component of the $\hat b$.

\subsection{Mode spectrum}
The kinetic energy is
\begin{align}
  T = \frac{1}{2} \dot{\mathbf{x}}^\top M \dot{\mathbf{x}}
  = \frac{1}{2} \dot{\mathbf{u}}^\top M \dot{\mathbf{u}},
\end{align}
where $M$ is the diagonal mass matrix.
Thus the equations of motion are
\begin{align}
  M \ddot{\mathbf u} = -K\mathbf{u} - \mathbf{L}
\end{align}
for a given external load $\mathbf{L}$. This can be derived from the Lagrangian
$\mathcal{L} = T - V - \mathbf{u}^\top\mathbf{L}$.

Using the ansatz $\mathbf{u} = e^{i\omega t} \mathbf{v}$, we obtain a generalized
eigenvector equation for the eigenmodes,
\begin{align}
  K \mathbf{v} = \omega^2 M \mathbf{v}.
\end{align}
Thus, the eigenmodes are given by the spectral decomposition of the
dynamical matrix
\begin{align}
    D = M^{-1} K.
\end{align}

\subsection{Springs with constant mass density}
Real springs have a mass, which is important when actually fabricating a
network. The potential energy is unchanged, but we get a new
kinetic term for the springs themselves.

Consider a rigid rod whose ends are displaced from
$\vec x_{1,2}^{(0)}$ to $\vec x_{1,2}$ by $\vec u_{1,2} = \vec x_{1,2} - \vec x_{1,2}^{(0)}$.
Parametrizing the points on the rod by $0 \leq s \leq 1$, a mass point
of mass $dm = m_s ds$ is displaced to $\vec x(s) = \vec x_1^{(0)}
+ \vec x_1 - \vec x_1^{(0)} + s(\vec x_2 - \vec x_1)$.
Thus, the whole rod has kinetic energy
\begin{align}
T_{\text{spring}} &= \frac{m_s}{2} \int_0^1 ds\, \dot{\vec{x}}(s)^2 \\
  &=   \frac{m_s}{2} \int_0^1 ds \left( \dot{\vec{u}}_1 + s\left(\dot{\vec{u}}_2
  - \dot{\vec{u}}_1 \right) \right)^2 \\
  &= \frac{m_s}{6}\left( \dot{\vec{u}}_1^2 + \dot{\vec{u}}_1^\top \dot{\vec{u}}_2 + \dot{\vec{u}}_2^2 \right).
  \label{eq:mass-matrix}
\end{align}
Therefore, the effect of springs with finite mass is the addition of
a non-diagonal effective mass matrix $S$ to the Lagrangian.

For springs with equal masses, the matrix $S$ has block
diagonal form $S = \tilde S^{\oplus d}$ with $d$ identical blocks,
one for each dimension.
The individual blocks have the form
\begin{align}
  \frac{12}{m_s} \tilde S_{ij} = \begin{cases}
  2\operatorname{deg}(i) & i=j \\
  1 & i\neq j \text{ and an edge connects $i$ and $j$} \\
  0 & \text{otherwise}
\end{cases}
\end{align}
$\tilde S$ is proportional to the matrix $2D + A$, where $D$ is the diagonal
degree matrix and $A$ is the adjacency matrix.
For springs with unequal masses, the adjacency matrix is weighted by
the spring masses and $D$ is similarly taken as the weighted degree matrix
obtained by taking the row or column sums of $A$.

Equivalently, taking $E$ as the oriented and $F$ as the non-oriented
incidence matrix of the network, we can write the manifestly positive
definite quadratic form
$\tilde S = (E m_s E^\top + 3 F m_s F^\top)/12$,
where $m_s$ is now the diagonal matrix of spring masses.
This is equivalent to decoupling the motion into translation of
the center-of-mass and a rigid rotation.

\subsection{Periodic crystals}
In periodic crystals, the components of the displacements $u_\sigma(\vec\ell)$ are indexed by node $\sigma$
and unit cell $\vec\ell$. In order to get around
the problem of finding the spectrum of an infinitely large stiffness
matrix, we Fourier transform away the periodic part of the displacements by
\begin{align}
  u_\sigma(\vec\ell) = \sum_{\vec q}
  e^{i \vec{q}^\top \vec{R}_{\sigma,\vec\ell}} u_\sigma(\vec q),
\end{align}
where the $\vec q$ run over vectors in the first Brillouin zone of the reciprocal
lattice and $\vec{R}_{\sigma,\vec\ell} = \vec{R}_{\vec\ell} + \vec{r}_\sigma$,
where $\vec{R}_{\vec\ell}$ is the
crystal lattice vector and $\vec{r}_\sigma$ the position vector of node
$\sigma$ in the unit cell.

By plugging this into the relation $\mathbf{e} = Q\mathbf{u}$
between elongations and displacements,
expanding $\mathbf{e}$ in Fourier modes as well, we find that
\begin{align}
  \mathbf{e}(\vec q) = Q(\vec q) \mathbf{u}(\vec q),
  \qquad Q_{\sigma,\beta}(\vec q) =
    \sum_{\vec\ell} e^{-i \vec{q}^\top (\vec{R}_{\sigma,{\vec\ell}} - \vec{R}_{\beta,0})}
    Q_{\sigma,\beta}({\vec\ell}, 0).
    \label{eq:ft}
\end{align}

Similarly, the elastic energy can be expressed as
\begin{align}
  V = \frac{1}{2N_c}\mathbf{u}(\vec q)^\dagger K(\vec q) \mathbf{u}(\vec q),
  \qquad K(\vec q) = Q(\vec q)kQ^\dagger(\vec q),
\end{align}
where $N_c$ is the number of unit cells in the crystal.
Thus, we have reduced the infinite dimensional problem to
many finite-dimensional problems parameterized by the wavenumber $\vec q$.
Ignoring spring masses for the moment and focusing on point masses $m \mathbb{1}$,
the kinetic term becomes $T = \frac{m}{2N_c} \mathbf{u}^\dagger(\vec q)
\mathbf{u}(\vec q)$, such that
we need to solve the eigenvalue problem
\begin{align}
  K(\vec q) \mathbf{u}(\vec q) = m \omega^2 \mathbf{u}(\vec q)
\end{align}
for all $\vec q$ in the first Brillouin zone.
For nontrivial spring masses (or more generally for non-diagonal mass matrices),
the mass matrix needs to be Fourier transformed as well, leading to the
generalized eigenvalue problem
\begin{align}
  K(\vec q) \mathbf{u}(\vec q) = \omega^2 M(\vec q) \mathbf{u}(\vec q).
\end{align}

Practically, the Fourier transform \eqref{eq:ft} is performed by
taking the equilibrium matrix $Q$ for the unit cell and multiplying
each entry $Q_{\sigma,\beta}$ by
$e^{-i \vec{q}^\top (\vec{R}_{\sigma,{\vec\ell}}(\beta) - \vec{R}_\beta)}$,
where we set
\begin{align}
  \vec{R}_{\sigma,{\vec\ell}}(\beta) &=
  \begin{cases}
    \vec{r}_\sigma, &\beta\text{ starts at }\sigma \\
    \vec{r}_\sigma + {\ell}_\beta \hat b, & \beta\text{ ends at }\sigma
  \end{cases} \\
  \vec{R}_\beta &= \vec{r}_\sigma + \frac{1}{2}\ell_\beta \hat b.
\end{align}
Here, $\ell_\beta$ is the length of bond $\beta$ and $\hat b$
is the unit vector pointing along $\beta$. Because the unit vector
is defined as in the previous section, it will point into the adjacent
unit cell for periodic edges.

\section{Response-Optimized networks at various gap positions}
\begin{figure}[h]
\includegraphics[height=.7\textwidth]{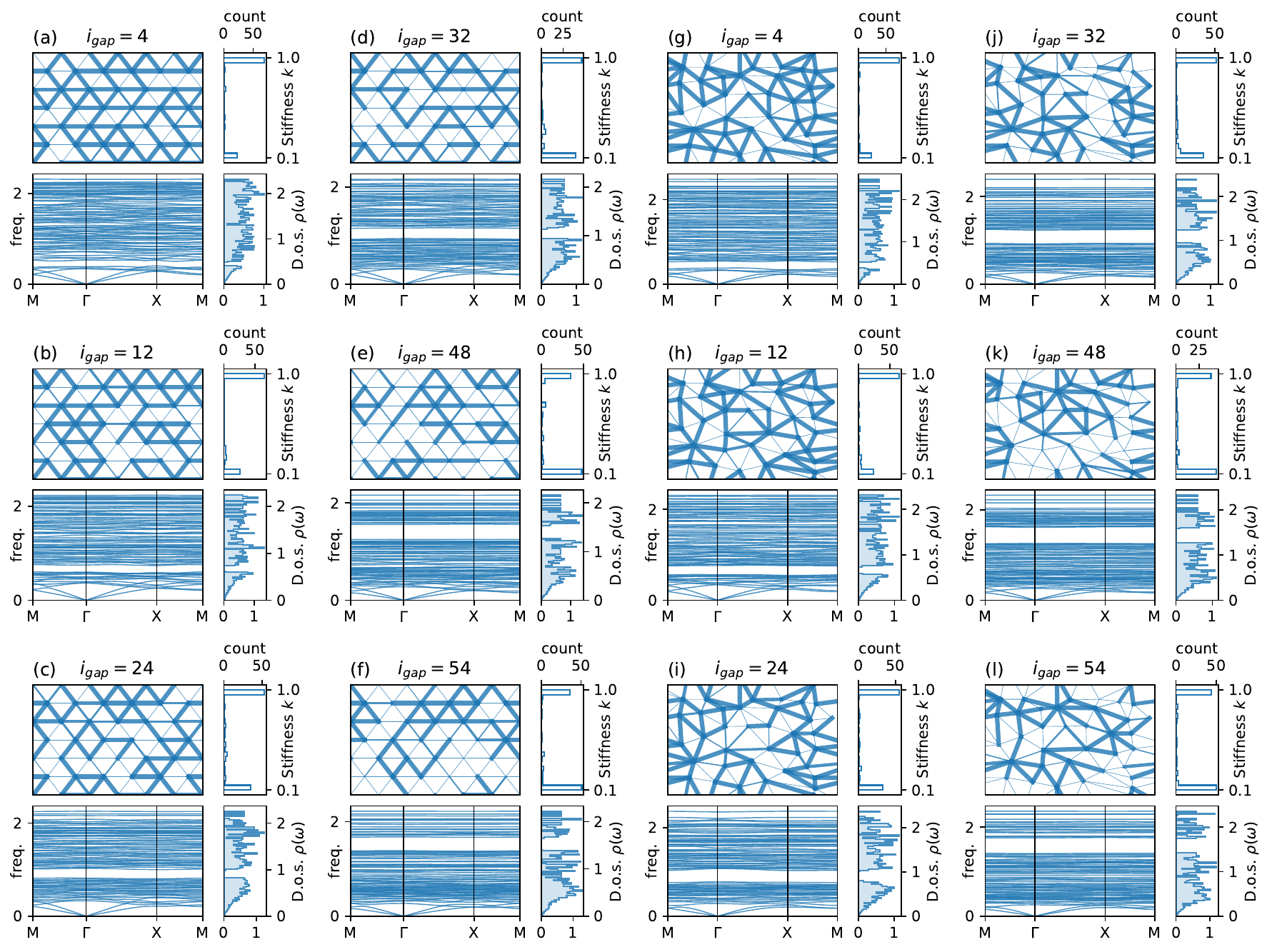}
\caption{Example networks with one gap
at index $i_{\mathrm{gap}}$ in a $6\times 6$
triangular grid \sbfig{a--f} and a randomized Delaunay network
with 36 nodes \sbfig{g--l}. In both cases, the higher the gap
position, the more weak bonds there are in the network
and the more disordered the stiffness pattern.
\label{fg:example-networks}}
\end{figure}
Here we show representative gap-optimized networks
with a single gap at various positions in the spectrum.
For both a triangular grid and a randomized
Delaunay network, the stiffness distribution is
concentrated at the bounds, and lower lying gaps
lead to a larger proportion of stiff bonds,
while higher lying gaps lead to a larger proportion
of weak bonds.
Results are shown in Fig.~\ref{fg:example-networks}.

\section{Networks of Elastic Rods}
\begin{figure}
    \centering
    \includegraphics[width=0.4\textwidth]{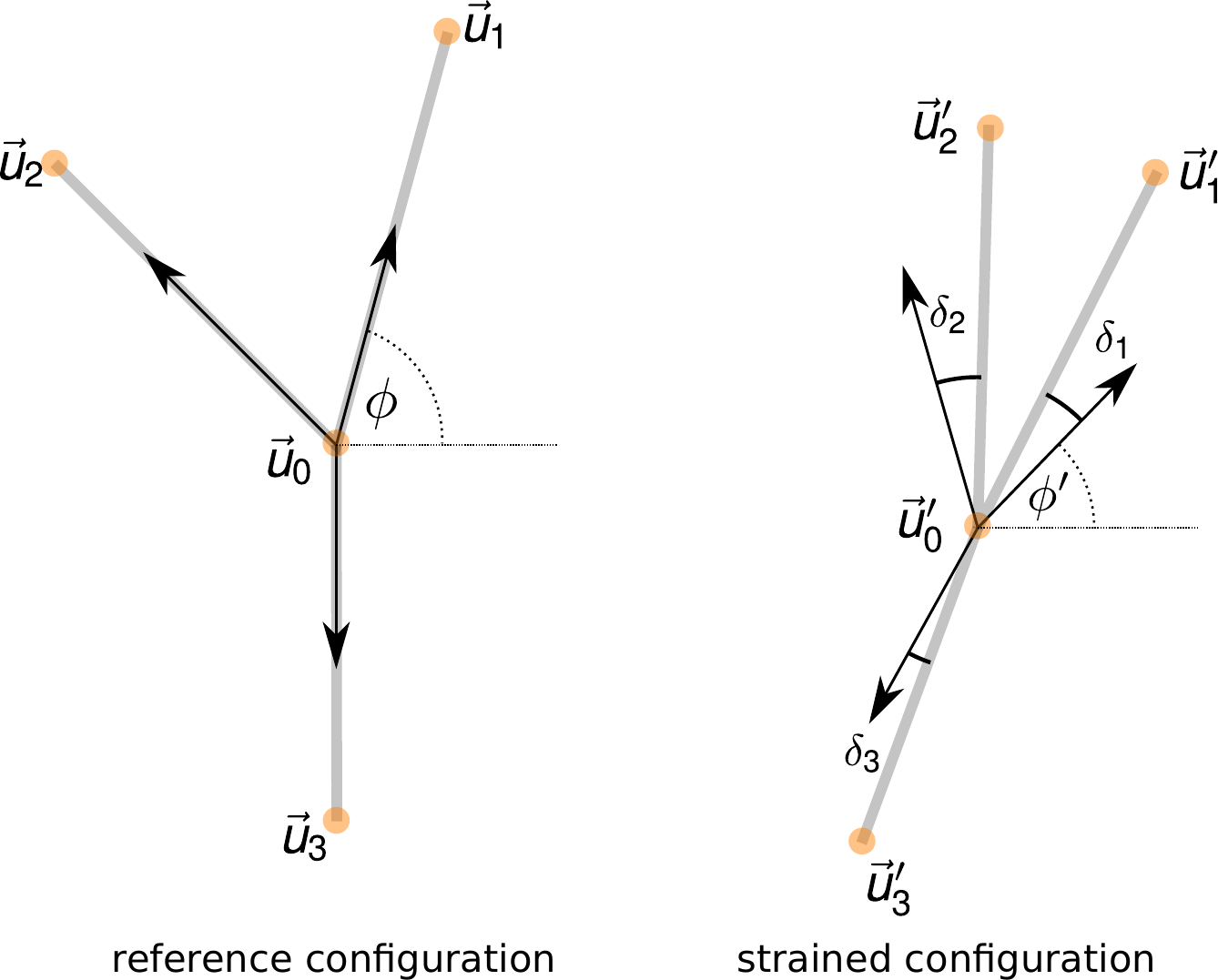}
    \caption{Rigid reference system attached to a node.
    In the reference configuration the rigid system carries an
    absolute orientation in space given by the angle
    $\phi$. In the strained configuration the node positions
    $\vec u_i$ have been changed to $\vec u_i'$. The absolute orientation
    of the reference system relaxes to a new value $\phi'$
    such that the angles $\delta_i$ between the elastic rods
    and their reference orientations are minimized.}
    \label{fig:reference_config}
\end{figure}
Here we present the model of locally preferred discrete reference coordinate
systems to model bond bending and stiff hinges.


A \emph{node} will be thought of as a point-like rigid body
itself, carrying a spatial orientation. Each one of the rods are then
attached to this rigid body at a particular orientation, and angular
deviations from this orientation will be penalized.
The orientation of the node itself then becomes a natural
part of the elastic energy, with the node orientation itself
adjusting to a minimal-energy configuration.

To get the idea, imagine a ball or a circle instead of a point-like node,
with the rods attached to the surface, and see Fig.~\ref{fig:reference_config}.

\subsection{Bending elastic energy}
In 2D, the rigid body at node $i$ is defined by vectors,
\begin{align}
    \vec b_{ij}^{(0)}
    = \begin{pmatrix}
    \cos(\delta_{ij}^{(0)} + \phi_i) \\
    \sin(\delta_{ij}^{(0)} + \phi_i)
    \end{pmatrix},
\end{align}
where the unit vectors point from node $i$ to node $j$,
the angles $\delta_{ij}$ define the reference configuration,
and $\phi_i$ defines the rigid orientation of node $i$.
Each rod $(ij)$ now tries to align to its reference configuration,
while the rigid node also rotates to adjust.
For simplicity, we consider the quadratic elastic energy at
a single node.
It now reads,
\begin{align}
    V_i = \kappa \sum_j \left( \delta_{ij} - \delta_{ij}^{(0)}
    + \phi_i \right)^2.
    \label{eq:energy}
\end{align}
In the quadratic case, we can directly solve for the
rigid part. The energy will be minimized, meaning that
\begin{align}
    0 &= \frac{\partial V_i}{\partial \phi_i} \\
    \Rightarrow 0 &= \sum_j \left( \delta_{ij} - \delta_{ij}^{(0)}
        + \phi_i \right) \\
    \Rightarrow \phi_i &= -\frac{1}{n} \sum_j \left(
         \delta_{ij} - \delta_{ij}^{(0)}  \right),
\end{align}
where $n$ is the number of rods attached to node $i$.
The elastic energy now reads,
\begin{align}
    V_i = \kappa \sum_j \left( \delta_{ij} - \delta_{ij}^{(0)}
     -\frac{1}{n} \sum_k \left(
         \delta_{ik} - \delta_{ik}^{(0)}  \right) \right)^2.
         \label{eq:quadratic-rigid}
\end{align}
Effectively, deviations from the mean deviation are penalized.
This is the standard deviation of angular deviations.
Let us write down the case of a discrete rod, a node with two
edges.
The elastic energy is
\begin{align}
    V_i &= \kappa  \left( \delta_{i+} - \delta_{i+}^{(0)}
     -\frac{1}{2} \left(
         \delta_{i+} - \delta_{i+}^{(0)} + \delta_{i-} - \delta_{i-}^{(0)} \right) \right)^2
        + \kappa  \left( \delta_{i-} - \delta_{i-}^{(0)}
     -\frac{1}{2} \left(
         \delta_{i+} - \delta_{i+}^{(0)} + \delta_{i-} - \delta_{i-}^{(0)} \right) \right)^2
         \\
       &= \frac{\kappa}{4}  \left( \delta_{i+} - \delta_{i+}^{(0)}
         - \delta_{i-} + \delta_{i-}^{(0)} \right)^2
        + \frac{\kappa}{4}  \left( \delta_{i-} - \delta_{i-}^{(0)}
         - \delta_{i+} + \delta_{i+}^{(0)} \right)^2 \\
       &= \frac{\kappa}{2}  \left( \delta_{i+}  - \delta_{i-} - \delta_{i+}^{(0)}
         + \delta_{i-}^{(0)} \right)^2.
\end{align}
The angular difference $\delta_{i+}  - \delta_{i-}$, where each angle
is measured with respect to, say the horizontal, is exactly the angle
between the two edges.
Thus, the elastic energy reduces to the standard discrete elastic rod case.

Alternatively, we can use the cosine energy,
\begin{align}
    V_i &= -\kappa \sum_j \vec b_{ij}^\top R(\phi_i) \vec b_{ij}^{(0)},
    \label{eq:cosine-rigid}
\end{align}
where $R(\phi) \in SO(2)$ is a rotation matrix.
We can still eliminate $\phi_i$ by,
\begin{align}
    0 &= \frac{\partial V_i}{\partial \phi_i} \\
    \Rightarrow 0 &= \sum_j \sin( \delta_{ij} - \delta_{ij}^{(0)}
        + \phi_i) \\
        &= \cos\phi_i \sum_j\sin(\delta_{ij} - \delta_{ij}^{(0)})
        + \sin\phi_i \sum_j \cos(\delta_{ij} - \delta_{ij}^{(0)}) \\
        \Rightarrow \tan\phi_i &= -\frac{\sum_j\sin(\delta_{ij} - \delta_{ij}^{(0)})}{
         \sum_j \cos(\delta_{ij} - \delta_{ij}^{(0)})},
\end{align}
which can be plugged into the formula for the rotation matrix $R(\phi_i)$.
In the case of small angular deviations, the cosine energy Eq.~\eqref{eq:cosine-rigid} reduces to the quadratic energy
Eq.~\eqref{eq:quadratic-rigid}, up to an irrelevant constant.

\section{Linearized dynamics}
In order to compute spectra and band structures, we need a way to
compute the dynamical matrix.
While Eq.~\eqref{eq:energy} is a simple quadratic in the angle,
it is highly nonlinear in the node positions that the angle
is constructed from. Therefore, linearization is necessary.

\subsection{Jacobian algebra}

We first compute the Jacobian of a unit vector $\vec b
= \vec x / |\vec x|$ with
respect to the coordinates, $J_{ij} = \partial b_i / \partial x_j$.
We find
\begin{align}
    J = \frac{1}{\ell} \begin{pmatrix}
    b_y^2 & -b_x b_y \\
    -b_x b_y & b_x^2
    \end{pmatrix} = \frac{1}{\ell} R\, \vec b \vec b^\top R^\top,
\end{align}
where $\ell = |\vec x|$ is the length of the vector that was
used to construct $\vec b$.
It is then also easy to verify the algebraic identities
\begin{align}
    J^2 = \frac{1}{\ell} J, \qquad R^\top J R = J = R J R^\top,
    \qquad J\, \vec b = 0, \qquad J R\, \vec b = \frac{1}{\ell} R\, \vec b.
\end{align}


\subsection{Dynamical matrix}
We now want to compute the new angles after a small change
$\vec x_i' = \vec x_i + \vec u_i$.
To linear order, we find that the unit vectors change as
\begin{align}
    \vec b_i' = \vec b_i + J_i \Delta\vec u_i,
\end{align}
where $\Delta \vec u_i = \vec u_i - \vec u_0$.

We consider the cosine energy first. For small angular differences
$\delta_{ij} - \delta^{(0)}_{ij}$, it is equivalent to the quadratic energy.
In this case, also the deviation from the reference orientation,
$\phi_i$ is small, and
\begin{align}
    \tan\phi_i \approx \phi_i &\approx -\frac{\sum_j (\delta_{ij} - \delta_{ij}^{(0)})}{\sum_j 1} \\
    \Rightarrow \phi_i &= -\frac{1}{n} \sum_j (\delta_{ij} - \delta_{ij}^{(0)}),
\end{align}
so we can focus on the quadratic energy Eq.~\eqref{eq:quadratic-rigid} entirely.
The dynamical matrix is then for the case where the mass of the node
is negligible (i.e., it always immediately relaxes to its equilibrium
while the rods vibrate on a longer timescale).

Expanding around the reference rest configuration it is sufficient to
find the lowest order contribution to the angular difference.
We easily obtain
\begin{align}
    \delta - \delta' &\approx \sin (\delta - \delta') \\
    &= \vec {b'}^\top R \vec b \\
    &= \underbrace{\vec b^\top\, R\, \vec b}_{=0} + \Delta \vec u^\top J R\, \vec b\\
    &= \frac{1}{\ell} (R\vec b)^\top \Delta\vec u \\
    &= \frac{1}{\ell} \vec d^\top \Delta\vec u,
\end{align}
where we introduced $\vec d = R\,\vec b = (b_y, -b_x)^\top$.
Then the elastic energy can be written to lowest order as
\begin{align}
    V_i &= \frac{\kappa}{2} \sum_j \left( \frac{1}{\ell_{ij}}\vec d_{ij}^\top
    \Delta\vec u_{ij} - \frac{1}{n_i} \sum_k  \frac{1}{\ell_{ik}}\vec d_{ik}^\top
    \Delta\vec u_{ik}  \right)^2 \\
    &= \frac{\kappa}{2} \sum_j \left(
        \sum_k \left( \delta_{kj} - \frac{1}{n_i} \right)
        \frac{1}{\ell_{ik}}\vec d_{ik}^\top \Delta\vec u_{ik}
        \right)^2.
    \label{eq:2d-linearized}
\end{align}
Now introduce the vector of flattened node positions $\vec u = (u_{1,x},
\dots, u_{N,x}, u_{1,y}, \dots, u_{N,y})^\top$ and the vector
$\vec D_{ik}$ such that $\vec D_{ik}^\top \vec u = \vec d_{ik}^\top
\Delta\vec u_{ik}/\ell_{ik}$.
These are the columns of the equilibrium
matrix, only constructed for rotated
bond vectors.
With this we further simplify the elastic energy,
\begin{align}
    V_i &= \frac{\kappa}{2} \sum_j \left(
        \sum_k \left( \delta_{kj} - \frac{1}{n_i} \right)
        \frac{1}{\ell_{ik}}\vec D_{ik}^\top \vec u
        \right)^2 \\
        &= \frac{\kappa}{2} \vec u^\top \left( \sum_{j,k,m}
        \left( \delta_{kj} - \frac{1}{n_i} \right) \left( \delta_{mj} - \frac{1}{n_i} \right) \vec D_{ik} \vec D_{im}^\top \right) \vec u \\
        &= \frac{1}{2} \vec u^\top D_i \, \vec u.
\end{align}
which is now entirely defined through the \emph{dynamical matrix} $D_i$.
In a system with multiple nodes, the total elastic energy is a sum over
that of all the individual nodes.
The total dynamical matrix can then be expressed as
\begin{align}
    D &= \sum_i D_i = \sum_{i,j,k,m} \kappa_i
        \left( \delta_{kj} - \frac{1}{n_i} \right) \left( \delta_{mj} - \frac{1}{n_i} \right) \vec D_{ik} \vec D_{im}^\top \\
        &= \sum_{i,j} \kappa_i \vec D_{ij} \vec D_{ij}^\top
        - \sum_i \kappa_i \frac{1}{n_i} \sum_{j,k} \vec D_{ij} \vec D_{ik}^\top \\
        &= \sum_{i,j} \kappa_i \vec D_{ij} \vec D_{ij}^\top
        - \sum_i \kappa_i \frac{1}{n_i} \vec D_i \vec D_i^\top,
        \label{eq:2d-dynamical-matrix}
\end{align}
where we introduced $\vec D_i = \sum_j \vec D_{ij}$.
The form of Eq.~\eqref{eq:2d-dynamical-matrix} explicitly
shows the character of the dynamics as a variance.
Even though it is not expressed as a sum of squares, it is
positive definite.

\section{Fitting elastic constants to FEM models}
\begin{figure}
    \centering
    \includegraphics[width=\textwidth]{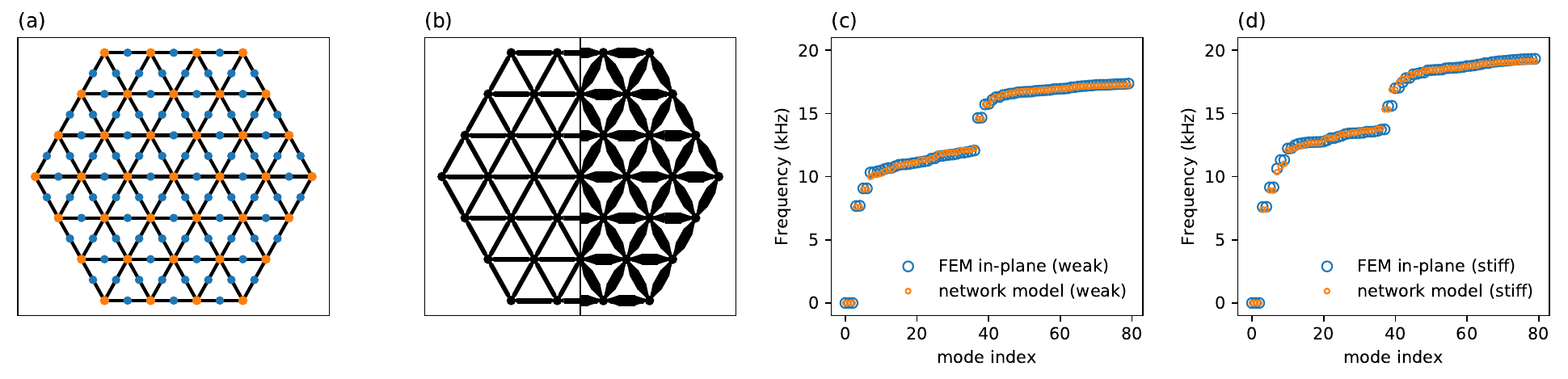}
    \caption{Fitting elastic constants to FEM simulated networks.
    \sbfig{a} Finite network topology modeling dynamics of continuum elastic networks. Orange nodes model stiff hinges
    with elastic constant $\kappa_h$, blue nodes model
    bending with elastic constant $\kappa_b$.
    \sbfig{b} 2D patterns for uniform weak (left) and stiff (right)
    continuum networks. Only half of each pattern is shown.
    The patterns are extruded in the $z$ direction
    (Fig.~3 if the main paper) and fed into MATLAB for finite element modal analysis.
    \sbfig{c} FEM modal analysis result for the
    first 80 mode frequencies in a uniform
    continuum elastic network with all weak bonds
    and fitted network model results.
    \sbfig{d} FEM modal analysis result for the
    first 80 mode frequencies in a uniform
    continuum elastic network with all stiff bonds
    and fitted network model results.
}
    \label{fg:fem-fit}
\end{figure}

In order to compare our network model with finite element calculations,
we calculate 80 FEM modes for continuum networks with the same topology
as Fig.~\ref{fg:fem-fit}~\sbfig{a} for the two cases where (i) all edges
are weak, and (i) all edges are
stiff (Fig.~\ref{fg:fem-fit}~\sbfig{b}).
The material constants are set for Styrodur (Materials
and Methods). The network diameter is approximately $17\,\mathrm{cm}$
and its height is $1\,\mathrm{cm}$.

This is fitted to a network model described by
\begin{align}
    m_s\, M \ddot{\mathbf{u}} + \left( k_s\, Q Q^\top + \kappa_b\, D_b
    + \kappa_h\, D_h  \right) \mathbf{u} = 0,
\end{align}
Where $m_s$ is the mass of the springs and $M$ is the mass matrix as defined
in Eq.~\eqref{eq:mass-matrix}, $k_s$ is the stretching stiffness of
the springs, $Q$ is the stretching compatibility matrix,
$\kappa_b$ is the bending stiffness, $D_b$ is the bending dynamical matrix
for the blue nodes in Fig.~\ref{fg:fem-fit}~\sbfig{a},
$\kappa_h$ is the hinge stiffness, and $D_h$ is the bending dynamical matrix
for the orange hinge nodes in Fig.~\ref{fg:fem-fit}~\sbfig{a}.

Because stretching modes contribute very little, we fix the stretching
stiffness to $k_s = 20$ for both weak and stiff networks.
This leaves the three fitting parameters $m_s$, $\kappa_b$, and $\kappa_h$.
The fits are performed by a non-linear least squares minimization on the
objective

\begin{align}
    \min_{m_s, \kappa_b,\kappa_h}\, \sum_{i=0}^{80} \left( \omega_{i}(m_s, \kappa_b,\kappa_h) - \hat\omega_{i,\text{FEM}} \right)^2,
\end{align}
where $\omega_{i}, \omega_{i,\text{FEM}}$ are the model and FEM eigenfrequencies,
respectively, and $\hat\omega_i = \omega_i/(20\, \mathrm{kHz})$ scales
the FEM eigenfrequencies to the interval $[0, 1]$.
In these dimensionless units the fit results are
\begin{center}
    \begin{tabular}{|c|c|c|}
    \hline
      &  weak & stiff \\
      \hline
      $m_s$ & $2.66$ & $2.93$ \\
      $\kappa_b$ & $0.10$ & $0.22$ \\
      $\kappa_h$ & $0.56$ & $0.68$ \\
      \hline
    \end{tabular}
\end{center}

For the spectral gap optimization we further simplify the model by
allowing only $\kappa_b$ as the optimization variable and
fixing $m_s=2.8$ and $\kappa_h=0.6$.

\section{In-plane and out-of-plane modes}
The 2D network approximation we use is justified because
the three-dimensional networks we employ have highly separated
in-plane ($x$--$y$) and out-of-plane ($z$) modes.
To demonstrate this, we separate the modes computed
using finite elements into
\begin{align}
    \mathbf{u} = (\mathbf{u}_x, \mathbf{u}_z, \mathbf{u}_z).
\end{align}
Then we compute the in-plane contribution
\begin{align}
    f_{xy}^2 = \frac{\|\mathbf{u}_x\|^2 + \|\mathbf{u}_y\|^2}{
    \|\mathbf{u}_x\|^2 + \|\mathbf{u}_y\|^2 + \|\mathbf{u}_z\|^2}.
\end{align}

\begin{figure}
    \centering
    \includegraphics[width=.35\textwidth]{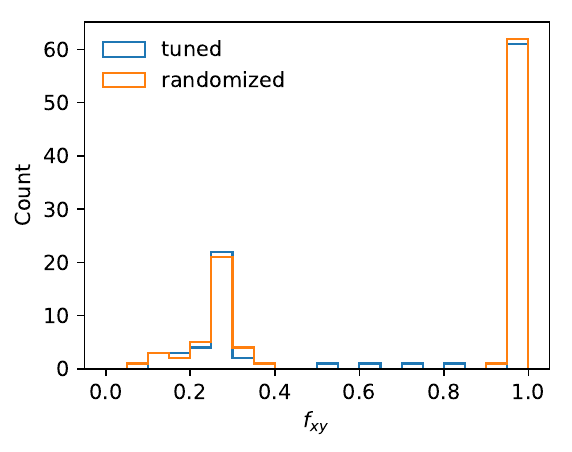}
    \caption{The histogram of in-plane contributions $f_{xy}$ shows
    that the first 100 modes are strongly separated for the tuned and randomized FEM networks
    from the main paper, justifying the 2D approximation for the in-plane
    modes.}
    \label{fg:in-plane}
\end{figure}

The histogram of the $f_{xy}$ for the tuned and the randomized
networks from the main paper computed from the first 100
FEM eigenmodes shows that the in-plane and the out-of-plane
responses are strongly separated, justifying our approach
(Fig.~\ref{fg:in-plane})

\section{Classical Brute-force optimization techniques}
\begin{figure}[h]
\includegraphics[width=0.6\textwidth]{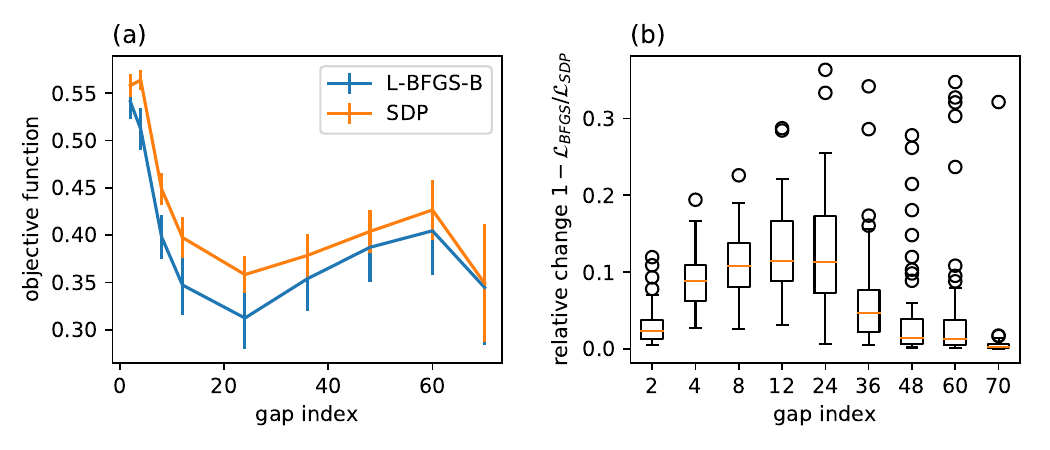}
\caption{Comparison of \texttt{L-BFGS-B} optimization algorithm
to the subspace method from Ref.~\cite{Men2010} in a $6\times 6$
periodic triangular grid for the objective function
Eq.~\eqref{eq:objective-gapsize} for gap index $i$.
Gap indices used were
$i = 2, 4, 8, 12, 24, 36, 48, 60, 70$, and optimizations were
carried out for 50 different random initial conditions at each gap
index.
(a) Mean value of the objective function for the derivative-free SDP algorithm
and the generic \texttt{L-BFGS-B} algorithm. Error bars correspond to
one standard deviation.
(b) Boxplots of the relative improvement that the SDP algorithm is able to achieve
over
the generic algorithm. While the SDP algorithm is always better, the improvement
is small (less than 5\%) for very low and medium to high lying gaps.
A large improvement (greater than 10\%)
is achieved only for gaps around index $i \approx 20$.
The median improvement exceeds 10\% only marginally at some gap
indices.
\label{fg:bfgs-sdp}}
\end{figure}

Here we show results of classical bandgap optimization techniques
to compare to our LRO method.
We consider both subspace optimization methods and naive
application of gradient based optimization to the
non-differentiable direct gap size objective.

The classical objective function for bandgap optimization
is~\cite{Men2010}
\begin{align}
 \mathcal{L} = \frac{\omega^2_{i+1} - \omega^2_i}{\omega^2_{i+1} + \omega^2_i},
 \label{eq:objective-gapsize}
\end{align}
maximizing the relative gap-midgap ratio of two consecutive
eigenvalues.

If the eigenstates directly above or below a band gap are
degenerate, the derivative
\begin{align}
 \frac{\partial (\omega^2_{i+1} - \omega^2_i)}{\partial k_e}
\end{align}
ceases to be well-defined, which can be immediately seen
by computing it from standard theory. This can be troublesome
for optimization algorithms which rely on the existence of these
derivatives, such as \texttt{L-BFGS-B}.
In order to circumvent this problem, methods have been developed
based on successive semidefinite programming that
do not rely on the existence of derivatives.
In the following, we compare a straightforward re-implementation
of the semidefinite programming (SDP) algorithm from Ref.~\cite{Men2010}
using the software packages CVXOPT and CVXPY to the results from \texttt{L-BFGS-B}
for various gap positions.
We chose to use up to 30\% of available states above and below the gap
and stopped the iterative algorithm when the relative change
in stiffnesses was less than $10^{-4}$.
We considered only the single wavevector $\mathbf{q} = 0$.
At each gap position we performed 10 optimizations from different,
random initial conditions using \texttt{L-BFGS-B}, the output of
which was then fed into the SDP algorithm as initial data.
Figure~\ref{fg:bfgs-sdp} compares the value of the objective function Eq.~\eqref{eq:objective-gapsize}
for the two algorithms. The networks used in the tests were periodic, triangular
grids with a unit cell size of $6\times 6$ nodes.

While the SDP algorithm is, unsurprisingly, always better than the
generic \texttt{L-BFGS-B} algorithm, the improvement is small
(less than 5\% on average) for very-low lying and medium-to-high lying gaps.
For gaps at low-to-medium positions, the SDP algorithm provides
a significant improvement (more than 10\% on average).
This, however, comes at the cost of significantly higher computation time.
For the networks considered, the SDP algorithm takes on the order of minutes,
while the generic algorithm takes on the order of a few seconds.

Figure~\ref{fg:figure1} directly compares the results
of optimizing the same initial conditions using the SDP and the naive
gradient based method.

Figures~\ref{fg:figure2} and \ref{fg:figure3} show further
results using the naive gradient based method only.

\begin{figure*}
  \includegraphics[width=.8\textwidth]{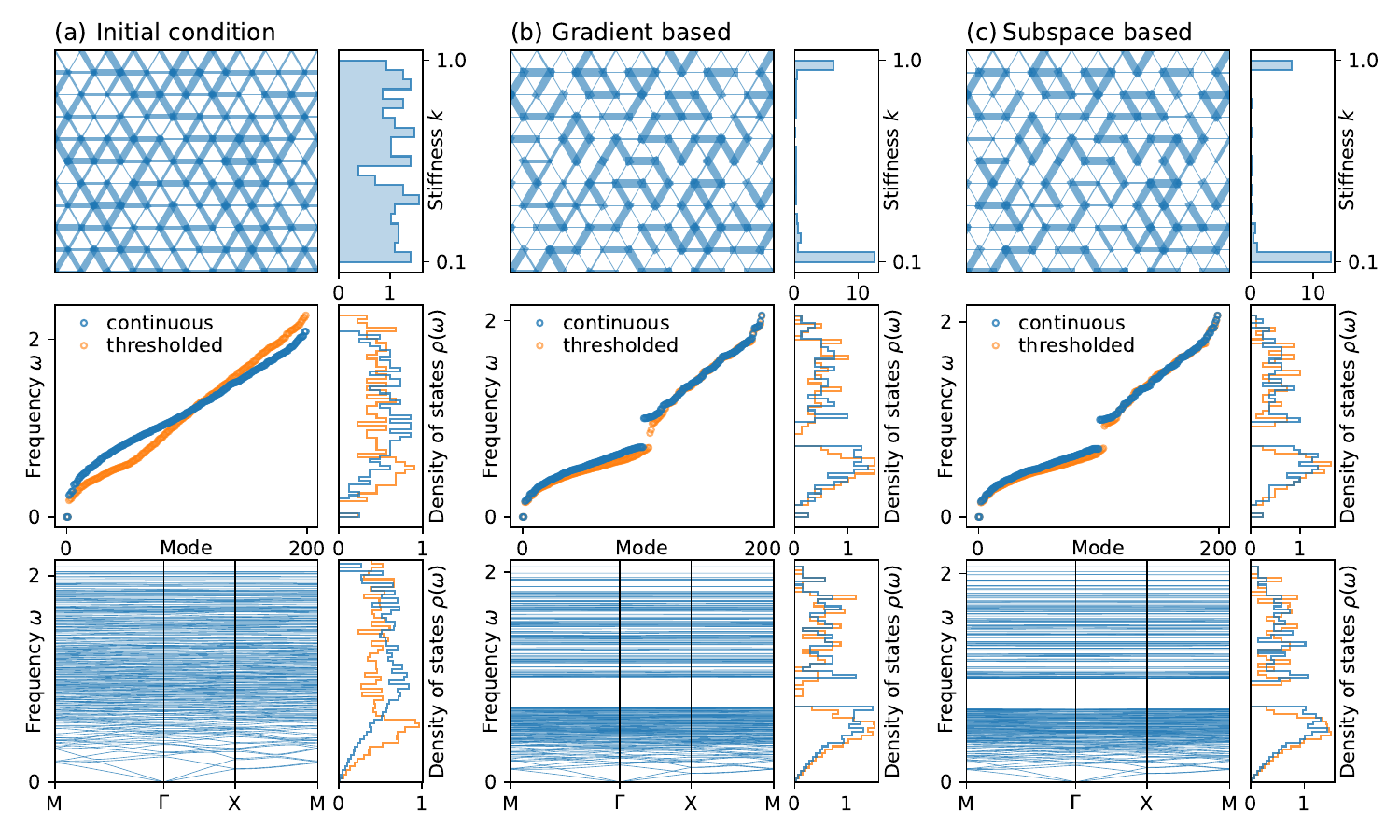}
  \caption{
  \textbf{Generating a bandgap in the unit cell of a 2D periodic
  mechanical network through direct numerical optimization}. The spring
  stiffnesses are bounded to the interval $[0.1, 1]$, and the target gap is imposed between
  modes $102$ and $103$ at wavevector $\mathbf{q} = \mathbf{0}$ only.
  The network is then thresholded by setting all
  $k \geq 0.55$ to $1$ and the remaining $k$ to $0.1$.
  \sbfig{a}~Initial conditions of the optimization. Top: The unit cell of a triangular grid
  of $10\times 10$ point masses $m$ connected by springs with uniformly random stiffnesses
  drawn from $(0.1, 1.0)$. Edge thicknesses are proportional to spring stiffness.
  Middle: The vibrational spectrum of the initial
  network and of the thresholded initial network.
  Bottom: The band structure of a 2D periodic crystal
  with the same rectangular unit cell and lattice vectors $\vec a=(10,0)$, $\vec b=(0,5\sqrt{3})$.
  The symmetry points in the Brillouin zone are
  $\Gamma = (0,0), X=(\pi/10,0), M=(\pi/10,\pi/(5\sqrt{3}))$.
  Both the spectrum of the unit cell and the band
  structure are dense and contain no gap.
  \sbfig{b}~Converged network after optimization using a gradient based algorithm
  (see section~Methods). The stiffnesses cluster
  tightly around the extremal values, and the gap is substantial in
  both the unit cell (middle) and the entire band structure (bottom). The thresholded network
  shows a spectral gap as well.
  \sbfig{c}~Refined network using the algorithm from Ref.~\cite{Men2010}
  with the output from \sbfig{b} as initial conditions.
  The gap in the thresholded network is now larger but the statistical
  features of the network remain almost identical.
  The objective function was improved from $0.335$ to $0.342$.
  \label{fg:figure1}}
\end{figure*}

\begin{figure*}[t]
  \includegraphics[width=.8\textwidth]{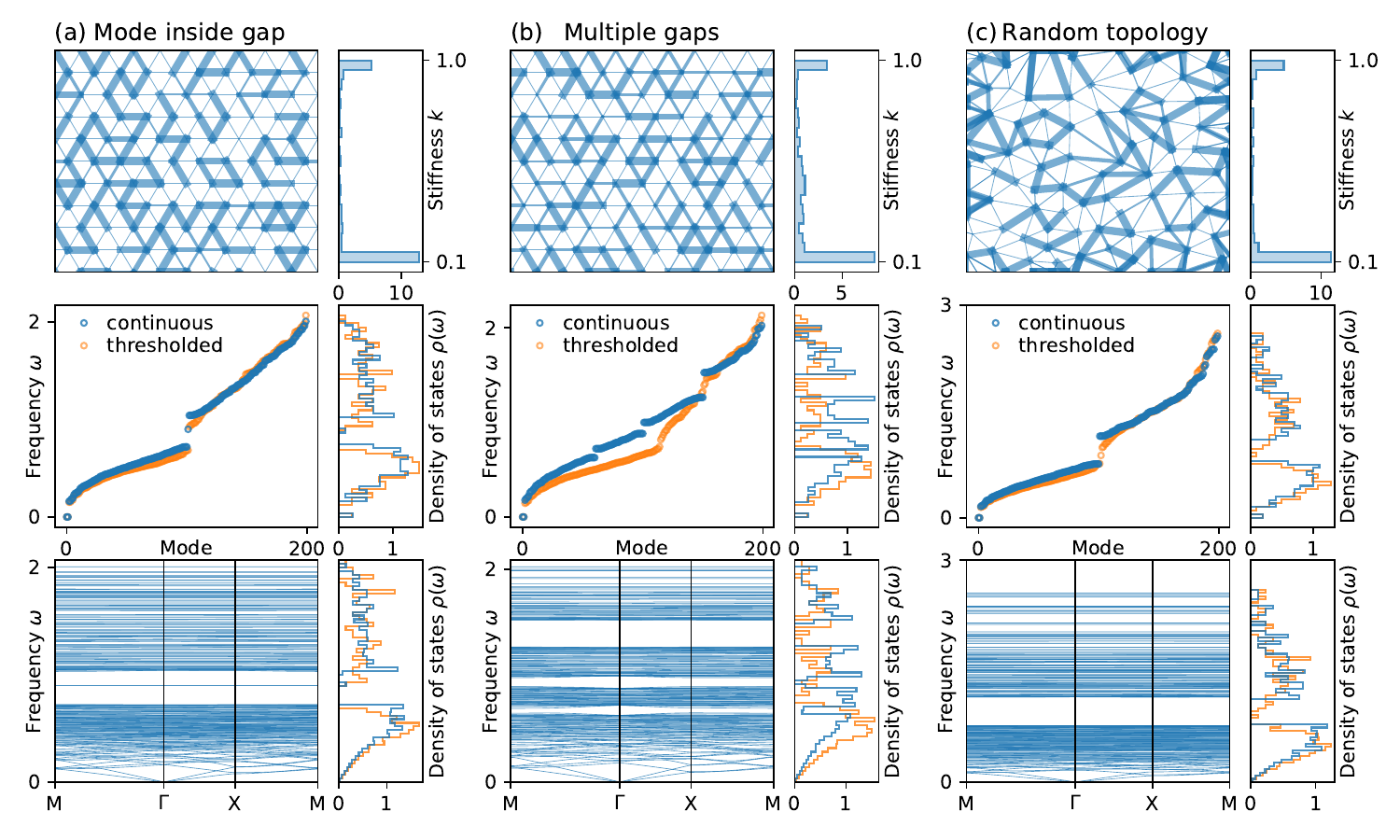}
  \caption{\textbf{Tuning more complex spectral properties and network topologies.}
  \sbfig{a}~Final state of a triangular lattice
  network with $10\times 10$ unit cell and imposed
  gaps between modes 100, 101, and 102, effectively leading to a single
  mode separated from the bulk spectrum.
  Despite tuning only at wavevector $\mathbf{q} = \mathbf{0}$, this property persists in the periodic lattice.
  \sbfig{b}~Final state of a triangular lattice
  network with $10\times 10$  unit cell and three imposed gaps
  at modes $60$, $100$, $150$. Again, the three gaps tuned into
  the unit cell spectrum survive into the periodic crystal.
  \sbfig{c}~Final state of a network with disordered unit cell consisting
  of $45$ point masses placed at random and their positions then
  relaxed (see section~Methods).
  The network was constructed by periodic Delaunay triangulation. The spectral gap survives into the band structure of the associated
  periodic crystal with square unit cell and high symmetry points
  $\Gamma = (0,0)$, $X=(\pi,0)$, $M=(\pi,\pi)$ in the Brillouin zone. Note that the thresholded networks in \sbfig{a}--\sbfig{c} do not exhibit the same properties as the continuous ones, suggesting that complex spectral characteristics are more fragile than simple gaps, and that networks with more than two different spring stiffnesses may be required to realize them.
  \label{fg:figure2}}
\end{figure*}

\begin{figure*}[t]
  \includegraphics[width=.8\textwidth]{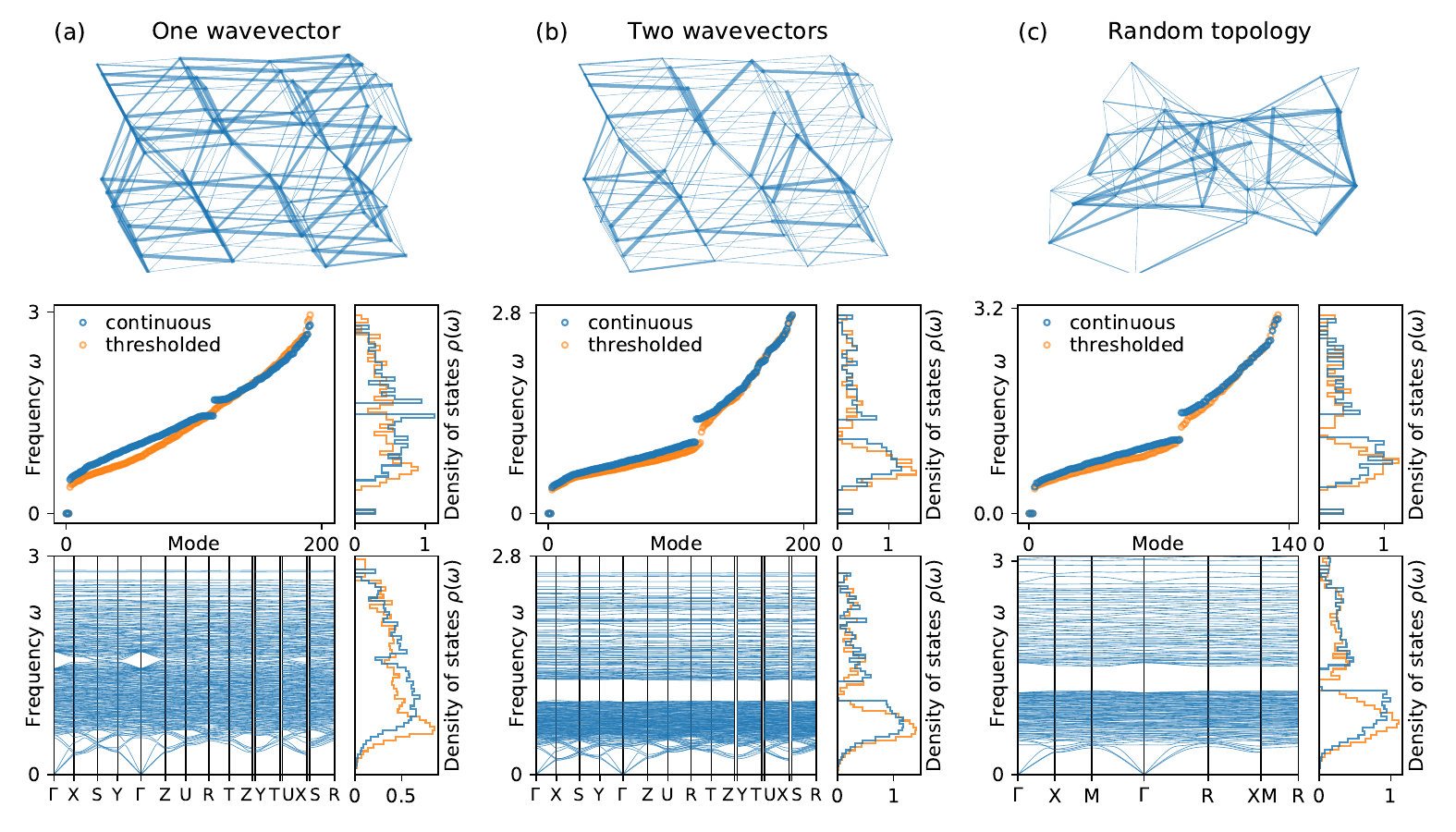}
  \caption{\textbf{Tuning bandgaps in 3D materials.}
  \sbfig{a}~$4\times 4\times 4$ unit cell of a periodic  tetrahedral network with a gap tuned between modes $102$ and $103$ (edges connecting adjacent unit cells not shown).  While the gap is large near the $\Gamma$ point of the associated
  orthorhombic crystal with lattice vectors $(4,0,0), (0,2\sqrt{3},0), (0,0,4\sqrt{2/3})$,
  the bands almost close in the rest of the Brillouin zone.
  \sbfig{b}~Unit cell of a similar tetrahedral network
  with a gap tuned
  between modes $102$ and $103$ using
  the summed objective function $\sum_{\{\mathbf{q}\}} \mathcal{L}(\mathbf{q})$ with
  $\{\mathbf{q}\}=\{{\bf 0}, (\frac{\pi}{8}, \tfrac{\pi}{4\sqrt{3}}, \tfrac{\pi\sqrt{3}}{8\sqrt{2}})\}$.
  Here, the spectral gap tuned into the spectra at only two points of the Brillouin zone remains open across the entire band structure.
  \sbfig{c}~Unit cell of a 3D network constructed from
  the periodic Delaunay triangulation of $45$ points with
  relaxed positions (cf. Fig.~\ref{fg:figure2}\sbfig{c})
  in the cubic unit cell $[0,1] \times [0,1] \times [0,1]$. Edges connecting adjacent unit cells not shown. A gap was tuned between
  modes $81$ and $82$ with the same objective function as in \sbfig{b} and $\{\mathbf{q}\}=\{{\bf 0}, (\pi/2,\pi/2,\pi/2)\}$.
  Again, by tuning the spectrum at only two points in the Brillouin zone, a gap spanning the entire band structure is  obtained. For all band structures, the symmetry points were defined as in~\cite{Setyawan2010}.
  \label{fg:figure3}}
\end{figure*}

\end{document}